\pdfoutput=1

\documentclass[12pt,english]{article}
\usepackage[T1]{fontenc}
\usepackage[latin9]{inputenc}

\usepackage{booktabs}

\newcommand{\mykeywords}{bang-bang controls, GMxB guarantees, convex
optimization, optimal stochastic control}

\usepackage{graphicx}
\usepackage{amsmath, amsthm, amssymb, amsfonts, latexsym}
\usepackage{array}
\usepackage{subfig}
\let\subfigure\subfloat
\usepackage[margin=1in]{geometry}
\usepackage{afterpage}

\usepackage{epsfig}
\usepackage{fancybox}
\usepackage{array}
\usepackage{shadow}

\usepackage{mathrsfs}
\usepackage{longtable}
\usepackage{calc}
\usepackage{multirow}
\usepackage{hhline}

\usepackage[running]{lineno}

\newcommand*\patchAmsMathEnvironmentForLineno[1]{%
  \expandafter\let\csname old#1\expandafter\endcsname\csname #1\endcsname
  \expandafter\let\csname oldend#1\expandafter\endcsname\csname end#1\endcsname
  \renewenvironment{#1}%
  {\linenomath\csname old#1\endcsname}%
  {\csname oldend#1\endcsname\endlinenomath}
}%
\newcommand*\patchBothAmsMathEnvironmentsForLineno[1]{%
  \patchAmsMathEnvironmentForLineno{#1}%
  \patchAmsMathEnvironmentForLineno{#1*}
}%
\AtBeginDocument{%
  \patchBothAmsMathEnvironmentsForLineno{equation}%
  \patchBothAmsMathEnvironmentsForLineno{align}%
  \patchBothAmsMathEnvironmentsForLineno{flalign}%
  \patchBothAmsMathEnvironmentsForLineno{alignat}%
  \patchBothAmsMathEnvironmentsForLineno{gather}%
  \patchBothAmsMathEnvironmentsForLineno{multline}%
}

\numberwithin{equation}{section}
\numberwithin{table}{section}
\numberwithin{figure}{section}

\newtheorem{definition}{Definition}
\newtheorem{theorem}[definition]{Theorem}
\newtheorem{lemma}[definition]{Lemma}
\newtheorem{remark}[definition]{Remark}

\newtheorem{corollary}[definition]{Corollary}

\newtheorem{example}[definition]{Example}

\numberwithin{definition}{section}

\newsavebox{\savepar}


\usepackage{lmodern}

\usepackage{microtype}

\usepackage{algorithm2e}

\usepackage{enumitem}

\usepackage{xcolor}
\definecolor{mylinkcolor}{HTML}{0066cc}
\definecolor{mycitecolor}{HTML}{008800}
\definecolor{myurlcolor}{HTML}{0066cc}
\usepackage[bottom]{footmisc}
\usepackage[numbers]{natbib}
\usepackage[pdftex]{hyperref}

\hypersetup{
  linkcolor=mylinkcolor,
  citecolor=mycitecolor,
  urlcolor=myurlcolor,
  colorlinks=true,
}
\providecommand{\doi}{}
\renewcommand{\doi}[1]{\href{https://doi.org/#1}{doi:
\begingroup\urlstyle{rm}\nolinkurl{#1}\endgroup}}

\usepackage{amstext}

\usepackage{verbatim}

\renewcommand{\leq}{\leqslant}
\renewcommand{\geq}{\geqslant}

\begin{document}

\title{The existence of optimal bang-bang\\controls for GMxB contracts\thanks{This work was supported by the Natural Sciences and
Engineering Research Council of Canada (NSERC) and by the Global Risk Institute
(Toronto).}}

\author{P. Azimzadeh\footnote{Cheriton School of Computer Science, University of
Waterloo, Waterloo ON, Canada N2L 3G1 \texttt{pazimzad[at]uwaterloo[dot]ca}}\and
P.A. Forsyth \footnote{Cheriton School of Computer Science, University of
Waterloo, Waterloo ON, Canada N2L 3G1 \texttt{paforsyt[at]uwaterloo[dot]ca}}}

\date{}

\maketitle

\begin{abstract}
\noindent

A large collection of financial contracts offering guaranteed minimum
benefits are often posed as control problems, in which at any point
in the solution domain, a control is able to take any one of an uncountable
number of values from the \emph{admissible set}. Often, such contracts
specify that the holder exert control at a finite number of deterministic
times. The existence of an \emph{optimal bang-bang control,} an
optimal control taking on only a finite subset of values from
the admissible set, is a common assumption in the literature. In
this case, the numerical complexity of searching for an optimal control
is considerably reduced. However, no rigorous treatment as to when
an optimal bang-bang control exists is present in the literature.
We provide the reader with a bang-bang principle from which the existence
of such a control can be established for contracts satisfying some
simple conditions. The bang-bang principle relies on the convexity
and monotonicity of the solution and is developed using basic results
in convex analysis and parabolic partial differential equations. We
show that a \emph{guaranteed lifelong withdrawal benefit} (GLWB)
contract admits an optimal bang-bang control. In particular, we find
that the holder of a GLWB can maximize a writer's losses by only ever
performing nonwithdrawal, withdrawal at exactly the contract rate,
or full surrender. We demonstrate that the related \emph{guaranteed
minimum withdrawal benefit} contract is not convexity preserving,
and hence does not satisfy the bang-bang principle other than in certain
degenerate cases.

\vspace{9pt}
\noindent
\textbf{Keywords:}
\mykeywords

\end{abstract}

\section{Introduction}

\subsection{Main results}

A large collection of financial contracts offering guaranteed minimum
benefits (GMxBs) are often posed as control problems \cite{bauer2008universal},
in which the control is able to take any one of an uncountable number
of values from the \emph{admissible set} at each point in its
domain. For example, a contract featuring regular withdrawals may
allow holders to withdraw any portion of their account. In the
following, we consider a control which maximizes losses for the writer
of the contract, hereafter referred to as an \emph{optimal control}.

A typical example is a \emph{guaranteed minimum withdrawal benefit}
(GMWB). If withdrawals are allowed at any time (i.e. ``continuously''),
then the pricing problem can be formulated as a singular control \cite{milevsky2006financial,dai2008guaranteed,huang:2010,huang2013analysis}
or an impulse control \cite{chen08a} problem.

In practice, the contract usually specifies that the control can only
be exercised at a finite number of deterministic \emph{exercise times}
$t_{0}<t_{1}<\cdots<t_{N-1}$ \cite{bauer2008universal,chen2008effect}.
The procedure for pricing such a contract using dynamic programming
proceeds backwards from the expiry time $t_{N}$ as follows:
\begin{enumerate}
\item Given the solution as $t\rightarrow t_{n+1}^{-}$, the solution as
$t\rightarrow t_{n}^{+}$ is acquired by solving an initial value
problem.
\item The solution as $t\rightarrow t_{n}^{-}$ is then determined by applying
an optimal control, which is found by considering a collection of
optimization problems.
\end{enumerate}
If, for example, a finite difference method is used to solve the
initial value problem from $t_{n+1}^{-}$ to $t_{n}^{+}$, an optimal
control is determined by solving an optimization problem at each grid
node, in order to advance the solution to $t_{n}^{-}$. Continuing
in this way, we determine the solution at the initial time.

If there exists an \emph{optimal bang-bang} control, an optimal
control taking on only a finite subset of values from the admissible
set, the numerical algorithm simplifies considerably. The existence
of such a control is a common assumption in insurance applications
\cite{bacinello2011variable,ngai2011longevity,holz2012gmwb}, although
no rigorous treatment is present in the literature. In this paper,
we will also consider a weaker condition, a {\em{bang-bang principle}}.
In this case, although an optimal control is not necessarily a finite
subset of values from the admissible set, we will see that a control
having this property can result in a large reduction in computational
complexity.

Our main result in this paper is the specification of sufficient conditions
which can be used to guarantee the existence
of an optimal bang-bang control. This result relies on
the convexity and monotonicity of the solution and follows from a
combination of basic results in convex analysis and parabolic partial
differential equations (PDEs). We demonstrate our results on two common
contracts in the GMxB family:
\begin{itemize}
\item The \emph{guaranteed lifelong withdrawal benefit} (GLWB) (a.k.a.
\emph{guaranteed minimum lifelong withdrawal benefits (GMLWB)}) admits
an optimal bang-bang control. In particular, we prove that a holder
can maximize the writer's losses by only ever performing

\begin{itemize}
\item nonwithdrawal,
\item withdrawal at the contract rate (i.e. never subject to a penalty), or
\item a full surrender (i.e. maximal withdrawal; may be subject to a penalty).
\end{itemize}
\item On the other hand, the \emph{guaranteed minimum withdrawal benefit}
(GMWB) is not necessarily convexity preserving, and does not  satisfy
a bang-bang principle other than in certain degenerate cases.
\end{itemize}
In the event that it is not possible to determine an optimal control
analytically, numerical methods are required. Standard techniques
in optimization are not always applicable, since these methods cannot
guarantee convergence to a global extremum. In particular, without
a priori knowledge about the objective functions appearing in the
family of optimization problems corresponding to optimal holder behavior
at the exercise times, a numerical method needs to resort to a linear
search over a discretization of the admissible set. Convergence to
a desired tolerance is achieved by refining this partition \cite{wang2008maximal}.
Only with this approach can we be assured of a convergent algorithm.
However, if an optimal bang-bang control exists, discretizing the
control set becomes unnecessary. Theoretically, this simplifies convergence
analysis. More importantly, in practice, this reduces the amount of
work per local optimization problem, often the bottleneck of any numerical
method.

\subsection{Insurance applications}

The GLWB is a response to a general reduction in the availability
of defined benefit pension plans \cite{butrica2009disappearing},
allowing the buyer to replicate the security of such a plan via a
substitute. The GLWB is bootstrapped via a lump sum payment $w_{0}$
to an insurer, which is invested in risky assets. We term this the
\emph{investment account}. Associated with the GLWB contract is the
\emph{guaranteed withdrawal benefit account}, referred to as the withdrawal
benefit for brevity. This account is also initially set to $w_{0}$.
At a finite set of deterministic \emph{withdrawal times}, the holder
is entitled to withdraw a predetermined fraction of the withdrawal
benefit (or any lesser amount), even if the investment account diminishes
to zero. This predetermined fraction is referred to as the \emph{contract
withdrawal rate}. If holders wish to withdraw in excess of the
contract withdrawal rate, they can do so upon the payment of a penalty.
Typical GLWB contracts include penalty rates that are decreasing functions
of time.

These contracts are often bundled with \emph{ratchets} (a.k.a. step-ups),
a contract feature that periodically increases the withdrawal benefit
to the investment account, provided that the latter has grown larger
than the former. Moreover, \emph{bonus} (a.k.a. roll-up) provisions
are also often present, in which the withdrawal benefit is increased
if the holder does not withdraw at a given withdrawal time. Upon death,
the holder's estate receives the entirety of the investment account.
We show that a holder can maximize the writer's costs by only ever
performing \emph{nonwithdrawal}, \emph{withdrawal at exactly the contract
rate}, or \emph{surrendering the entirety of their account}. Such
a holder will never withdraw a nonzero amount strictly below the contract
rate or perform a partial surrender. However, this result requires
a special form for the penalty and lapsation functions, which is not
universal in all contracts. Pricing GLWB contracts has previously been
considered in \cite{piscopo2011valuation,holz2012gmwb,forsyth2013risk,azimzadeh2013hedging}.

Much like the GLWB contract, a GMWB is composed of an investment account
and withdrawal benefit initially set to $w_{0}$, in which $w_{0}$
is a lump sum payment to an insurer. At a finite set of withdrawal
times, the holder is entitled to withdraw up to a predetermined amount.
Note that this amount is not a fraction of the withdrawal benefit,
as in the GLWB, but rather a constant amount irrespective of the withdrawal
benefit's size. Furthermore, unlike the GLWB, the action of withdrawing
decreases both the investment account and withdrawal benefit on a
dollar-for-dollar basis.

The GMWB promises to return at least the entire original investment,
regardless of the performance of the underlying risky investment.
The holder may withdraw more than the predetermined amount subject
to a penalty. Upon death, the contract is simply transferred to the
holder's estate, and hence mortality risk need not be considered.
Pricing GMWB contracts has been previously considered in \cite{milevsky2006financial,dai2008guaranteed,chen2008effect,huang:2010,huang2013analysis}.

\subsection{Overview}

In $\S$\ref{sec:GMxBs}, we introduce the GLWB and GMWB contracts.
In $\S$\ref{sec:Model}, we generalize this to model a contract that
can be controlled at finitely many times, a typical case in insurance
practice (e.g. yearly or quarterly exercise). In $\S$\ref{sec:ControlReduction},
we develop sufficient conditions for the existence of an optimal bang-bang
control and show that the GLWB satisfies these conditions. $\S$\ref{sec:NumericalResults}
discusses a numerical method for finding the cost of funding GLWB
and GMWB contracts, demonstrating the bang-bang principle for the
former and providing an example of where it fails for the latter.

\section{Guaranteed minimum benefits (GMxBs)\label{sec:GMxBs}}

We introduce mathematical models for the GLWB and GMWB contracts in
this section.
Since most GMxB contracts offer withdrawals
on anniversary dates, to simplify notation, we restrict our attention
to annual withdrawals occurring at
\[
\mathscr{T}\equiv\left\{ 0,1,\ldots,N-1\right\} .
\]
$0$ and $N$ are referred to as the \emph{initial} and \emph{expiry}
times, respectively (no withdrawal occurs at $N$).

In order to ensure that the writer can, at least in theory, hedge
a short position in a GMxB with no risk, we assume that the holder
will employ a loss-maximizing strategy. That is, the holder will act
so as to maximize the cost of funding the GMxB. This represents the
worst-case hedging cost for the writer. This worst-case cost is a
function of the holder's investment account and withdrawal benefit.
As such, we write $\mathbf{x}\equiv(x_{1},x_{2})$, where
$x_{1}$ is the value of the investment account and $x_{2}$ is the
value of the withdrawal benefit. Both of these quantities are nonnegative.

Let $\alpha$ denote the \emph{hedging fee}, the rate continuously
deducted from the investment account $X_{1}$ (while $x_{1}$ is used
to denote a particular value of the investment account, the capital
symbol $X_{1}$ is reserved for the corresponding stochastic process)
to provide the premium for the contract. We assume that between exercise
times, the investment account of the GMxBs follows geometric Brownian
motion (GBM) as per
\[
\frac{dX_{1}}{X_{1}}=\left(\mu-\alpha\right)dt+\sigma dZ
\]
tracking the index $\hat{X}_{1}$ satisfying
\[
\frac{d\hat{X}_{1}}{\hat{X}_{1}}=\mu dt+\sigma dZ
\]
where $Z$ is a Wiener process under the \emph{real-world measure}.
We assume that it is not possible to short the investment account
$X_{1}$ for fiduciary reasons \cite{chen2008effect}, so that the
obvious arbitrage opportunity is prohibited.

The worst-case cost of a GMxB is posed as the solution to an initial
value problem (IVP) specified by three conditions:
\begin{enumerate}
\item the worst-case cost of funding the contract at the expiry time (posed
as a Cauchy boundary condition; see, for example, (\ref{eq:GMWB-Initial})
and (\ref{eq:GLWB-Initial2}));
\item the evolution of the worst-case cost \emph{across} withdrawals (posed
as a supremum over the holder's actions, corresponding to the holder
acting so as to maximize the writer's losses; see, for example, (\ref{eq:GMWB-Supremum})
and (\ref{eq:GLWB-Supremum}));
\item the evolution of the worst-case cost \emph{between} withdrawals (posed
as a conditional expectation; see, for example, (\ref{eq:GMWB-ValueFunction})
and (\ref{eq:GLWB-ValueFunction})).
\end{enumerate}
We begin by introducing the IVP for the GMWB before moving to the
GLWB for ease of exposition.
To distinguish the two contracts, we use the superscripts
$\text{L}$ and $\text{M}$ to denote quantities that pertain to the
GLWB and GMWB, respectively.
In the following, we denote by $\tilde{\mathbb{E}}$
the expectation and by $\tilde{Z}$ a Wiener process under the \emph{risk-neutral
measure, }that which renders the discounted index $\hat{X}_{1}$ into
a martingale. For a function $g$ whose domain is a subset of $\mathbb{R}$,
we use the notations $g(t^{-})\equiv\lim_{s\uparrow t}g(s)$
and $g(t^{+})\equiv\lim_{s\downarrow t}g(s)$
to denote the one-sided limits at $t$.

\subsection{Guaranteed minimum withdrawal benefit (GMWB)}

Since the GMWB is transferred to the holder's estate upon death, mortality
risk is not considered. The worst-case cost of funding a GMWB at time
$N$ (the expiry) is \cite{dai2008guaranteed}
\[
\varphi^{\text{M}}(\mathbf{x})\equiv\max(x_{1},\left(1-\kappa_N\right)x_{2}),
\]
corresponding to the greater of the entirety of the investment account
or a full surrender at the \emph{penalty rate} at the $N$th anniversary,
  $\kappa_N \in [0,1]$. The worst-case cost of funding
a GMWB at previous times is derived by a hedging argument in which the writer
takes a position in the index $\hat{X}_{1}$ \cite{chen2008effect}. Equivalently,
it is given by finding $V$ (within the relevant space of functions;
see Appendix \ref{app:Preservation}) such that (s.t.)
\begin{align}
V(\mathbf{x},N) & =\varphi^{\text{M}}(\mathbf{x}) & \text{on }\left[0,\infty\right)^{2}\label{eq:GMWB-Initial}\\
V(\mathbf{x},n^{-}) & =\sup_{\lambda\in[0,1]}\left[V(\mathbf{f}_{\mathbf{x},n}^{\text{M}}(\lambda),n^{+})+f_{\mathbf{x},n}^{\text{M}}(\lambda)\right] & \text{on }\left[0,\infty\right)^{2}\times\mathscr{T}\label{eq:GMWB-Supremum}\\
V(\mathbf{x},t) & =\tilde{\mathbb{E}}\Bigl[e^{-\int_{t}^{n+1}r(\tau)d\tau}V(X_{1}(\left(n+1\right)^{-}),x_{2},\left(n+1\right)^{-})\nonumber \\
  & \hspace{15.15em}\Big|
      X_{1}(t)=x_{1}\Bigr] & \text{ on }
        \left[0,\infty\right)^{2}\times\left(n,n+1\right)\,\forall n\label{eq:GMWB-ValueFunction}
\end{align}
where between exercise times
\begin{equation}
\frac{dX_{1}}{X_{1}}=\left(r-\alpha\right)dt+\sigma d\tilde{Z}. \label{eq:GMWB-XRiskNeutral}
\end{equation}
$r$ is the risk-free rate,
$f^{\text{M}}:[0,1]\rightarrow\mathbb{R}$
represents the cash flow from the writer to the holder, and $\mathbf{f}^{\text{M}}:[0,1]\rightarrow[0,\infty)^{2}$
represents the state of the contract postwithdrawal.  The construction
of $f^{\text{M}}$ and $\mathbf{f}^{\text{M}}$ is outlined below.
The holder is able to withdraw a fraction $\lambda\in[0,1]$
of the withdrawal benefit at each exercise time.

Intuitively, $V(\mathbf{x},n^{-})$ and $V(\mathbf{x},n^{+})$
can be thought of as the value of the contract ``immediately before''
and ``immediately after'' the exercise time $n$.

Let $G\geq0$ denote the \emph{predetermined contract withdrawal amount}
associated with the GMWB so that $G\wedge x_{2}$ ($a\wedge b\equiv\min(a,b)$,
$a\vee b\equiv\max(a,b)$) is the maximum the holder can
withdraw without incurring a penalty (both $\wedge$ and $\vee$ are
understood to have lower operator precedence than the arithmetic operations).
Consider the point $(x_{1},x_{2},n)$ with $n\in\mathscr{T}$.
\begin{itemize}
\item The maximum a holder can withdraw without incurring a penalty is $G\wedge x_{2}$.
If the holder withdraws the amount $\lambda x_{2}$ with $\lambda x_{2}\in[0,G\wedge x_{2}]$,
\begin{equation}
V(\mathbf{x},n^{-})=V(\underbrace{x_{1}-\lambda x_{2}\vee0,\, x_{2}-\lambda x_{2}}_{\mathbf{f}^{\text{M}}},\, n^{+})+\underbrace{\lambda x_{2}}_{f^{\text{M}}}.\label{eq:GMWB-Withdrawal}
\end{equation}

\item Let $\kappa_{n}\in[0,1]$ denote the \emph{penalty rate}
at the $n$th anniversary. If the holder withdraws the amount
$\lambda x_{2}$ with $\lambda x_{2}\in(G\wedge x_{2},x_{2}]$,
\begin{equation}
V(\mathbf{x},n^{-})
   =V(
     \underbrace{x_{1}-\lambda x_{2} \vee0,\, x_{2}-\lambda x_{2}}_{\mathbf{f}^{\text{M}}},\, n^{+})+\underbrace{\lambda x_{2}-\kappa_{n}\left(\lambda x_{2}-G\right)}_{f^{\text{M}}}.\label{eq:GMWB-Surrender}
\end{equation}
Here, $\lambda x_{2}\in(G\wedge x_{2},x_{2})$ corresponds
to a partial surrender and $\lambda x_{2}=x_{2}$ (i.e. $\lambda=1$)
corresponds to a full surrender.
\end{itemize}
We can summarize (\ref{eq:GMWB-Withdrawal}) and (\ref{eq:GMWB-Surrender})
by taking
\begin{equation}
f_{\mathbf{x},n}^{\text{M}}(\lambda)\equiv\begin{cases}
\lambda x_{2} & \text{if }\lambda x_{2}\in\left[0,G\wedge x_{2}\right]\\
G+\left(1-\kappa_{n}\right)\left(\lambda x_{2}-G\right) & \text{if }\lambda x_{2}\in\left(G\wedge x_{2},x_{2}\right]
\end{cases}\label{eq:GMWB-CashFlow}
\end{equation}
and
\[
\mathbf{f}_{\mathbf{x},n}^{\text{M}}(\lambda)\equiv\left( x_{1}-\lambda x_{2}\vee0,\left(1-\lambda\right)x_{2}\right) .
\]

It can be shown from (\ref{eq:GMWB-ValueFunction}) that the cost
to fund the GMWB (between exercise times) satisfies%
\footnote{We discuss what it means for a function to satisfy this PDE in Appendix
\ref{app:Preservation}.%
} \cite{chen2008effect}
\begin{equation}
\partial_{t}V+\mathcal{L}V=0\text{ on }\left(0,\infty\right)^{2}\times\left(n,n+1\right) \text{ } \forall n\label{eq:GMWB-PDE}
\end{equation}
 where
\begin{equation}
\mathcal{L}\equiv\frac{1}{2}\sigma^{2}x_{1}^{2}\partial_{x_{1}x_{1}}+\left(r-\alpha\right)x_{1}\partial_{x_{1}}-r.\label{eq:GMWB-L}
\end{equation}

\subsection{Guaranteed lifelong withdrawal benefit (GLWB)}

Let $\mathcal{M}(t)$ be the mortality rate at time $t$
(i.e. $\int_{t_{1}}^{t_{2}}\mathcal{M}(t)dt$ is the fraction
of the original holders who pass away in the interval $[t_{1},t_{2}]$),
so that the survival probability at time $t$ is
\[
\mathcal{R}(t)=1-\int_{0}^{t}\mathcal{M}(s)ds.
\]
We assume $\mathcal{M}$ is continuous and nonnegative, along with
$\mathcal{R}(t)\geq0$ for all times $t$.  We assume
that mortality risk is diversifiable. Furthermore, we assume the existence
of a time $t^{\star}>0$ s.t. $\mathcal{R}(t^{\star})=0$.
That is, survival beyond $t^{\star}$ is impossible (i.e. no holder
lives forever). $N$ is chosen s.t. $N\geq t^{\star}$ to ensure that
all holders have passed away at the expiry of the contract. As is
often the case in practice, we assume ratchets are prescribed to occur
on a subset of the anniversary dates (e.g. triennially).

As usual, we assume that the holder of a GLWB will employ a loss-maximizing
strategy. Since $N$ was picked sufficiently large, the insurer has
no obligations at the $N$th anniversary and the worst-case
cost of funding a GLWB at time $N$ is
\begin{equation}
\varphi^{\text{L}}(\mathbf{x})\equiv0.\label{eq:GLWB-Initial}
\end{equation}
As with the GMWB, the worst-case cost of funding a GLWB is derived
by a hedging argument in which the writer takes a position in the
index $\hat{X}_{1}$ \cite{forsyth2013risk}. Equivalently, it is
given by finding $V$ (within the relevant space of functions; see
Appendix \ref{app:Preservation}) s.t.
\begin{align}
V(\mathbf{x},N) & =\varphi^{\text{L}}(\mathbf{x}) & \text{on }\left[0,\infty\right)^{2}\label{eq:GLWB-Initial2}\\
V(\mathbf{x},n^{-}) & =\sup_{\lambda\in[0,2]}\left[V(\mathbf{f}_{\mathbf{x},n}^{\text{L}}(\lambda),n^{+})+f_{\mathbf{x},n}^{\text{L}}(\lambda)\right] & \text{on }\left[0,\infty\right)^{2}\times\mathscr{T}\label{eq:GLWB-Supremum}\\
V(\mathbf{x},t) & =\tilde{\mathbb{E}}\Bigl[e^{-\int_{t}^{n+1}r(\tau)d\tau}V(X_{1}(\left(n+1\right)^{-}),x_{2},\left(n+1\right)^{-})\nonumber \\
 & \qquad+\int_{t}^{n+1}e^{-\int_{t}^{s}r(\tau)d\tau}\mathcal{M}(s)X_{1}(s)ds\Big|
             X_{1}(t)=x_{1}\Bigr] & \text{ on }\left[0,\infty\right)^{2}\times\left(n,n+1\right)\,\forall n\label{eq:GLWB-ValueFunction}
\end{align}
where between exercise times, $X_{1}$ is specified by (\ref{eq:GMWB-XRiskNeutral}).

$f^{\text{L}}:[0,2]\rightarrow\mathbb{R}$ represents the
(mortality-adjusted \cite{forsyth2013risk}) cash flow from the writer
to the holder and $\mathbf{f}^{\text{L}}\colon[0,2]\rightarrow[0,\infty)^{2}$
represents the state of the contract postwithdrawal. In particular,
$\lambda=0$ corresponds to nonwithdrawal, $\lambda\in(0,1]$
corresponds to withdrawal at or below the contract rate, and $\lambda\in(1,2]$
corresponds to a partial or full surrender.

\begin{remark}We remark that the admissible set of actions $[0,2]$
is undesirably large (i.e. a continuum). We will apply the results
established in $\S$\emph{\ref{sec:ControlReduction}} to show that an optimal
strategy taking on values only from $\{ 0,1,2\} $ exists.
In other words, an equivalent problem can be constructed by substituting
the set $\{ 0,1,2\} $ for the original $[0,2]$
in the optimization problem \emph{(\ref{eq:GLWB-Supremum})}. The resulting
problem has smaller computational complexity than the original one
(i.e. successive refinements of $[0,2]$ need not be considered
to attain convergence).\end{remark}

The construction of $f^{\text{L}}$ and $\mathbf{f}^{\text{L}}$ is
guided by the specification of the contract:
\begin{itemize}
\item Let $\beta\geq0$ denote the \emph{bonus rate}: if the holder does not
withdraw, the withdrawal account is amplified by $1+\beta$.
\item Let $\delta\in[0,1]$ denote the \emph{contract withdrawal rate}; that is,
$\delta x_{2}$ is the maximum a holder can withdraw without incurring
a penalty.
\item Let $\kappa_{n}\in[0,1]$ denote the \emph{penalty rate}
at the $n$th anniversary, incurred if the holder withdraws
above the contract withdrawal rate.
\item Let
\[
\mathbb{I}_{n}=\begin{cases}
1 & \text{\text{if a ratchet is prescribed to occur on the }}n\text{th anniversary}\\
0 & \text{otherwise}
\end{cases}.
\]

\end{itemize}
Then,
\begin{equation}
f_{\mathbf{x},n}^{\text{L}}(\lambda)\equiv\mathcal{R}(n)\cdot\begin{cases}
0 & \text{if }\lambda=0\\
\lambda\delta x_{2} & \text{if }\lambda\in\left(0,1\right]\\
\delta x_{2}+\left(\lambda-1\right)\left(1-\kappa_{n}\right)\left(x_{1}-\delta x_{2}\vee0\right) & \text{if }\lambda\in\left(1,2\right]
\end{cases}\label{eq:GLWB-CashFlow}
\end{equation}
and
\begin{equation}
\mathbf{f}_{\mathbf{x},n}^{\text{L}}(\lambda)\equiv\begin{cases}
\left( x_{1},\, x_{2}\left(1+\beta\right)\vee\mathbb{I}_{n}x_{1}\right)  & \text{if }\lambda=0\\
\left( x_{1}-\lambda\delta x_{2}\vee0,\, x_{2}\vee\mathbb{I}_{n}\left[x_{1}-\lambda\delta x_{2}\right]\right)  & \text{if }\lambda\in\left(0,1\right]\\
\left(2-\lambda\right)\mathbf{f}_{\mathbf{x},n}^{\text{L}}(1) & \text{if }\lambda\in\left(1,2\right]
\end{cases}.\label{eq:GLWB-StateTransition}
\end{equation}

It can be shown that the cost to fund the GLWB (between exercise times)
satisfies \cite{forsyth2013risk}
\begin{equation}
\partial_{t}V+\mathcal{L}V+\mathcal{M}x_{1}=0\text{ on }\left(0,\infty\right)^{2}\times\left(n,n+1\right) \text{ } \forall n\label{eq:GLWB-PDE}
\end{equation}
where $\mathcal{L}$ is defined in (\ref{eq:GMWB-L}).

\section{General formulation\label{sec:Model}}

We generalize now the above IVPs. Let $\mathscr{T}\equiv\{ t_{0},\ldots,t_{N-1}\} $
along with the order $0\equiv t_{0}<\cdots<t_{N}\equiv T$, in which
$T$ is referred to as the expiry time. Let $\Omega$ be a convex
subset of $\mathbb{R}^{m}$. The set of all actions a holder can perform
at an exercise time $t_{n}$ is denoted by $\Lambda_{n}\subset\mathbb{R}^{m^{\prime}}$,
assumed to be nonempty, and referred to as an \emph{admissible set}.
For brevity, let
\begin{equation}
v_{\mathbf{x},n}(\lambda)\equiv V(\mathbf{f}_{\mathbf{x},n}(\lambda),t_{n}^{+})+f_{\mathbf{x},n}(\lambda)\label{eq:Model-LittleV}
\end{equation}
where $f_{\mathbf{x},n}\colon\Lambda_{n}\rightarrow\mathbb{R}$ and
$\mathbf{f}_{\mathbf{x},n}\colon\Lambda_{n}\rightarrow\Omega$. We
write $v_{\mathbf{x},n}(\lambda)$ to stress that for
each fixed $(\mathbf{x},n)$, we consider an optimization
problem in the variable $\lambda$. The general problem is to find
$V$ satisfying the conditions
\begin{align}
V(\mathbf{x},T) & =\varphi(\mathbf{x}) & \text{on }\Omega\label{eq:Model-Initial}\\
V(\mathbf{x},t_{n}^{-}) & =\sup v_{\mathbf{x},n}(\Lambda_{n}) & \text{on }\Omega\times\mathscr{T}\label{eq:Model-Supremum}
\end{align}
along with a condition specifying the evolution of $V$ from $t_{n}^{+}$
to $t_{n+1}^{-}$ (see, for example, (\ref{eq:GMWB-ValueFunction}) and
(\ref{eq:GLWB-ValueFunction})).

\begin{remark}\label{Model-AdmissibleSetIndependentOfX}Convexity
preservation, a property that helps establish the bang-bang principle,
depends on each admissible set $\Lambda_{n}$ being independent of
the state of the contract, $\mathbf{x}$. This is discussed in Remark
\emph{\ref{rmk:ControlReduction-Statelessness}}.\end{remark}

\section{Control reduction}\label{sec:ControlReduction}
\begin{definition}[Optimal bang-bang control]
\label{def:OptimalBangBang}$V$, a solution to the general IVP introduced
in $\S$\ref{sec:Model}, is said to admit an optimal bang-bang control
at time $t_{n}\in\mathscr{T}$ whenever
\[
V(\mathbf{x},t_{n}^{-})=\max v_{\mathbf{x},n}(\hat{\Lambda}_{n})\text{on }\Omega,
\]
where $\hat{\Lambda}_{n}$ denotes a finite set
independent of $\mathbf{x}$.
\end{definition}

The above condition is inherently simpler than (\ref{eq:Model-Supremum}),
in which there are no guarantees on the cardinality of $\Lambda_{n}$.

$\S$\ref{sub:ControlReduction-BangBangPrinciple} develops Corollary
\ref{cor:OptimalBangBang}, establishing sufficient conditions for
the existence of an optimal bang-bang control. This result requires
that the relevant solution $V$ be convex and monotone (CM). Given
a CM initial condition (\ref{eq:Model-Initial}), we seek to ensure
that $V$ preserves the CM property at all previous times. $\S$\ref{sub:ControlReduction-PreservationAcross}
develops conditions on the functions $f$ and $\mathbf{f}$ to ensure
that the supremum (\ref{eq:Model-Supremum}) preserves the CM property.
Similarly, $\S$\ref{sub:ControlReduction-Between} develops conditions
on the dynamics of $V$ (and hence the underlying stochastic process(es))
to ensure that the CM property is preserved between exercise times.

For the remainder of this work, we use the shorthand $V_{n}^{+}(\mathbf{x})\equiv V(\mathbf{x},t_{n}^{+})$
and $V_{n}^{-}(\mathbf{x})\equiv V(\mathbf{x},t_{n}^{-})$.

\subsection{Preliminaries}

In an effort to remain self-contained, we provide the reader with
several elementary (but useful) definitions. In practice, we consider only
vector spaces over $\mathbb{R}$ and hence restrict our definitions
to this case.
\begin{definition}[convex set]
Let $W$ be a vector space over  $\mathbb{R}$. $X\subset W$ is
a convex set if for all $x,x^{\prime}\in X$ and $\theta\in(0,1)$,
$\theta x+(1-\theta)x^{\prime}\in X$.
\end{definition}

\begin{definition}[convex function]
\label{def:ControlReduction-Convex}Let $X$ be a convex set and
$Y$ be a vector space over  $\mathbb{R}$ equipped with a partial
order $\leq_{Y}$. $h\colon X\rightarrow Y$ is a convex function
if for all $x,x^{\prime}\in X$ and $\theta\in(0,1)$,
\[
h(\theta x+\left(1-\theta\right)x^{\prime})\leq_{Y}\theta h(x)+\left(1-\theta\right)h(x^{\prime}).
\]

\end{definition}

\begin{definition}[extreme point]
An extreme point of a convex set $X$ is a point $x\in X$ which
cannot be written $x=\theta x^{\prime}+(1-\theta)x^{\prime\prime}$
for any $\theta\in(0,1)$ and $x^{\prime},x^{\prime\prime}\in X$
with $x^{\prime}\neq x^{\prime\prime}$.
\end{definition}

\begin{definition}[convex polytope]
Let $Y$ be a topological vector space over  $\mathbb{R}$. $P\subset Y$
is a convex polytope if it is a compact convex set with finitely many
extreme points. The extreme points of a convex polytope are referred
to as its vertices.
\end{definition}

\begin{definition}[monotone function]
\label{def:ControlReduction-Monotone}Let $X$ and $Y$ be sets equipped
with partial orders $\leq_{X}$ and $\leq_{Y}$, respectively. $h\colon X\rightarrow Y$
is monotone if for all $x,x^{\prime}\in X$, $x\leq_{X}x^{\prime}$
implies $h(x)\leq_{Y}h(x^{\prime})$.\end{definition}
\begin{lemma}
\label{lem:ControlReduction-ConvexComposition}Let $A$ be a convex
set, and let $B$ and $C$ be vector spaces over  $\mathbb{R}$ equipped
with partial orders $\leq_{B}$ and $\leq_{C}$, respectively. If
$h_{1}\colon A\rightarrow B$ and $h_{2}\colon B\rightarrow C$ are
convex functions with $h_{2}$ monotone, then $h_{2}\circ h_{1}$
is a convex function.
\end{lemma}
\begin{remark}\label{rmk:ControlReduction-OrderOnRN}For the remainder
of this work, we equip $\mathbb{R}^{m}$ with the order $\leq$ defined
as follows: if $\mathbf{x},\mathbf{y}\in\mathbb{R}^{m}$, $\mathbf{x}\leq\mathbf{y}$
whenever $x_{i}\leq y_{i}$ for all $i$.\end{remark}

\subsection{Bang-bang principle\label{sub:ControlReduction-BangBangPrinciple}}

Consider a particular exercise time $t_n$. Suppose the following:
\begin{enumerate}[label=(A\arabic*)]
\item \label{itm:ControlReduction-VConvexAndMonotone} $\mathbf{x} \mapsto V_{n}^{+}(\mathbf{x})$ is
CM. 
\item \label{itm:ControlReduction-BoundedFromAbove}For each fixed $\mathbf{x}\in\Omega$,
$v_{\mathbf{x},n}(\Lambda_n)$ is bounded above. 
\end{enumerate}
Throughout this section, we consider a particular point $\mathbf{y}\in\Omega$
in order to establish our result pointwise. For the results below, we require the following propositions:
\begin{enumerate}[label=(B\arabic*)]
\item \label{itm:ControlReduction-ConvexCollection}There exists a collection
$\mathcal{P}_{n}(\mathbf{y})\subset2^{\Lambda_{n}}$ s.t.
$\bigcup_{P\in\mathcal{P}_{n}(\mathbf{y})}P=\Lambda_{n}$
and each $P\in\mathcal{P}_{n}(\mathbf{y})$ is compact
convex.
\item \label{itm:ControlReduction-FlowAndStateConvexity}For each $P\in\mathcal{P}_{n}(\mathbf{y})$,
the restrictions $\lambda \mapsto f_{\mathbf{y},n}|_{P}(\lambda)$ and $\lambda \mapsto \mathbf{f}_{\mathbf{y},n}|_{P}(\lambda)$
are convex.
\item \label{itm:ControlReduction-FinitePolytopes}$\mathcal{P}_{n}(\mathbf{y})$
is a finite collection of convex polytopes.
\end{enumerate}
\begin{remark}\emph{\ref{itm:ControlReduction-ConvexCollection}} simply
states that we can ``cut up'' the admissible set $\Lambda_{n}$ into
(possibly overlapping) compact convex sets.
\emph{\ref{itm:ControlReduction-FlowAndStateConvexity}}
states that the restrictions of $f_{\mathbf{y},n}$ and $\mathbf{f}_{\mathbf{y},n}$
on each of these sets are convex functions of $\lambda$.\end{remark}
\begin{lemma}
\label{lma:ControlReduction-LittleVConvex}Suppose \emph{\ref{itm:ControlReduction-VConvexAndMonotone}},
\emph{\ref{itm:ControlReduction-ConvexCollection}}, and \emph{\ref{itm:ControlReduction-FlowAndStateConvexity}}.
For each $P\in\mathcal{P}_{n}(\mathbf{y})$, the restriction
$\lambda \mapsto v_{\mathbf{y},n}|_{P}(\lambda)$ is convex.\end{lemma} 
\begin{proof}
The proof is by (\ref{eq:Model-LittleV}), \ref{itm:ControlReduction-VConvexAndMonotone},
\ref{itm:ControlReduction-FlowAndStateConvexity}, and Lemma \ref{lem:ControlReduction-ConvexComposition}.\end{proof}
\begin{lemma}
\label{lem:ControlReduction-LittleVSupremum}Suppose \emph{\ref{itm:ControlReduction-VConvexAndMonotone}},
\emph{\ref{itm:ControlReduction-BoundedFromAbove}}, \emph{\ref{itm:ControlReduction-ConvexCollection}},
and \emph{\ref{itm:ControlReduction-FlowAndStateConvexity}}. Let $P\in\mathcal{P}_{n}(\mathbf{y})$ be nonempty.
Then,
\[
\sup v_{\mathbf{y},n}(P)=\sup v_{\mathbf{y},n}(E(P))
\]
where $E(P)$ denotes the set of extreme points of $P$.\end{lemma}
\begin{proof}
Let $w\equiv v_{\mathbf{y},n}|_{P}$. Note that $w(P)=v_{\mathbf{y},n}(P)$,
and hence no generality is lost in considering $w$. Lemma \ref{lma:ControlReduction-LittleVConvex}
establishes the convexity of $w$. Naturally, $\sup w(P)$
exists (and hence $\sup w(E(P))$ exists too)
due to \ref{itm:ControlReduction-BoundedFromAbove}. Finally, it is
well known from elementary convex analysis that the supremum of a
convex function on a compact convex set $P$ lies on the extreme points
of $P$, $E(P)$. See \cite[Sec.~32]{rockafellar1997convex}.
\end{proof}

\begin{theorem}[bang-bang principle]
\label{thm:ControlReduction-SupremumEverywhere}Suppose \emph{\ref{itm:ControlReduction-VConvexAndMonotone}},
\emph{\ref{itm:ControlReduction-BoundedFromAbove}}, \emph{\ref{itm:ControlReduction-ConvexCollection}},
and \emph{\ref{itm:ControlReduction-FlowAndStateConvexity}}.  Then,
\[
\sup v_{\mathbf{y},n}(\Lambda_{n})=\sup v_{\mathbf{y},n}\Biggl(\bigcup_{P\in\mathcal{P}_{n}(\mathbf{y})}E(P)\Biggr)
\]
where $E(P)$ denotes the set of extreme points of $P$.\end{theorem}
\begin{proof}
By \ref{itm:ControlReduction-ConvexCollection}, we have that $\Lambda_{n}=\bigcup_{P\in\mathcal{P}_{n}(\mathbf{y})}P$.
We can, w.l.o.g., assume that all members of $\mathcal{P}_{n}(\mathbf{y})$
are nonempty (otherwise, remove all empty sets). $\sup v_{\mathbf{y},n}(\Lambda_{n})$
exists due to \ref{itm:ControlReduction-BoundedFromAbove}. Since
for each $P\in\mathcal{P}_{n}(\mathbf{y})$, $\sup v_{\mathbf{y},n}(P)=\sup v_{\mathbf{y},n}(E(P))$
(Lemma \ref{lem:ControlReduction-LittleVSupremum}), two applications of Lemma \ref{lem:Commutativity-ResultNonEmpty} allow us to ``commute'' the supremum with the union to get
\[
\sup v_{\mathbf{y},n}(\Lambda_{n}) =\sup v_{\mathbf{y},n}\Biggl(\bigcup_{P\in\mathcal{P}_{n}(\mathbf{y})}P\Biggr)
 =\sup v_{\mathbf{y},n} \Biggl(\bigcup_{P\in\mathcal{P}_{n}(\mathbf{y})}E(P)\Biggr).\qedhere
\]

\end{proof}

Theorem \ref{thm:ControlReduction-SupremumEverywhere} reduces the
region over which to search for an optimal control. When $\mathcal{P}_{n}(\mathbf{y})$
is a finite collection of convex polytopes, the situation is even nicer,
as $\bigcup_{P\in\mathcal{P}_{n}(\mathbf{y})}E(P)$
is a finite set (a finite union of finite sets). If, in addition, $\mathcal{P}_{n}$
is chosen independent of $\mathbf{y}$, we arrive at an optimal bang-bang
control:
\begin{corollary}[optimal bang-bang control]
\label{cor:OptimalBangBang}Suppose \emph{\ref{itm:ControlReduction-VConvexAndMonotone}} and
\emph{\ref{itm:ControlReduction-BoundedFromAbove}}. Furthermore, suppose \emph{\ref{itm:ControlReduction-ConvexCollection}},
\emph{\ref{itm:ControlReduction-FlowAndStateConvexity}}, and \emph{\ref{itm:ControlReduction-FinitePolytopes}}
for all $\mathbf{y}\in\Omega$. Finally, suppose that there exists
$\mathcal{P}_{n}$ s.t. $\mathcal{P}_{n}=\mathcal{P}_{n}(\mathbf{y})$
for all $\mathbf{y}\in\Omega$. Then, the general IVP introduced in
$\S$\ref{sec:Model} admits an optimal bang-bang control at time
$t_{n}$ (Definition \emph{\ref{def:OptimalBangBang}}) with
\[
V(\mathbf{x},t_{n}^{-})=\sup v_{\mathbf{x},n}(\Lambda_{n})=\max v_{\mathbf{x},n}(\hat{\Lambda}_{n})\text{ on }\Omega
\]
and
\[
\hat{\Lambda}_{n}\equiv\bigcup_{P\in\mathcal{P}_{n}}E(P).
\]

\end{corollary}
\begin{example}\label{exm:ControlReduction-GLWBBangBang}Let $\mathbf{y}\in[0,\infty)^{2}$.
We now find $\mathcal{P}_{n}^{\text{L}}(\mathbf{y})$ s.t.
\emph{\ref{itm:ControlReduction-ConvexCollection}}, \emph{\ref{itm:ControlReduction-FlowAndStateConvexity}},
and \emph{\ref{itm:ControlReduction-FinitePolytopes}} are satisfied for
the GLWB. Take $P_{1}\equiv[0,1]$, $P_{2}\equiv[1,2]$,
and $\mathcal{P}_{n}^{\text{L}}(\mathbf{y})\equiv\{ P_{1},P_{2}\} $,
satisfying \emph{\ref{itm:ControlReduction-FinitePolytopes}}. Note that
$\bigcup_{P\in\mathcal{P}_{n}^{\text{L}}(\mathbf{y})}P=[0,2]$,
satisfying \emph{\ref{itm:ControlReduction-ConvexCollection}}. It is trivial
to show that the functions $f_{\mathbf{y},n}^{\text{L}}|_{P_{j}}$
and $\mathbf{f}_{\mathbf{y},n}^{\text{L}}|_{P_{j}}$ defined in \emph{(\ref{eq:GLWB-CashFlow})}
and \emph{(\ref{eq:GLWB-StateTransition})} are convex as functions of $\lambda$
(the maximum of convex functions is a convex function), thereby satisfying
\emph{\ref{itm:ControlReduction-FlowAndStateConvexity}}. Since $\mathbf{y}$
was arbitrary and $\mathcal{P}_{n}^{\text{L}}$ was chosen independent
of $\mathbf{y}$, we conclude (whenever \emph{\ref{itm:ControlReduction-VConvexAndMonotone}}
and \emph{\ref{itm:ControlReduction-BoundedFromAbove}} hold), by Corollary
\emph{\ref{cor:OptimalBangBang}}, that the supremum of $v_{\mathbf{y},n}^{\text{L}}$
occurs at
\[
\hat{\Lambda}_{n}^{\text{L}}=E(P_{1})\cup E(P_{2})=E(\left[0,1\right])\cup E(\left[1,2\right])=\left\{ 0,1\right\} \cup\left\{ 1,2\right\} =\left\{ 0,1,2\right\}
\]
(corresponding to nonwithdrawal, withdrawal at exactly the contract
rate, and a full surrender).\end{example}

\begin{remark}
When all the conditions required for Corollary
\emph{\ref{cor:OptimalBangBang}} hold,
with the exception that $\mathcal{P}_{n}(\mathbf{y})$
depends on $\mathbf{y}$, then an optimal control is
not necessarily bang-bang, but does satisfy the bang-bang principle,
Theorem \emph{\ref{thm:ControlReduction-SupremumEverywhere}}.
This may still result in considerable
control-reduction.

\end{remark}

\subsection{Preservation of convexity and monotonicity across exercise times\label{sub:ControlReduction-PreservationAcross}}

Since the convexity and monotonicity of $V$ are desirable properties
upon which the bang-bang principle depends (i.e. \ref{itm:ControlReduction-VConvexAndMonotone}),
we would like to ensure that they are preserved ``across'' exercise
times (i.e. from $t_{n}^{+}$ to $t_{n}^{-}$).

Consider the $n$th exercise time, $t_{n}$. Suppose the following:

\begin{enumerate}[label=(C\arabic*)]
\item \label{ass:ControlReduction-FlowAndStateConvexityInX}For each fixed
$\lambda\in\Lambda_{n}$, $\mathbf{x}\mapsto\mathbf{f}_{\mathbf{x},n}(\lambda)$
and $\mathbf{x}\mapsto f_{\mathbf{x},n}(\lambda)$ are convex.%
\footnote{Note that this is not the same as \ref{itm:ControlReduction-FlowAndStateConvexity}
Here, we mean that for each fixed $\lambda\in\Lambda_{n}$ and for
all $\mathbf{x},\mathbf{x}^{\prime}\in\Omega$ and $\theta\in(0,1)$,
\[
f_{\theta\mathbf{x}+(1-\theta)\mathbf{x}^{\prime},n}(\lambda)\leq\theta f_{\mathbf{x},n}(\lambda)+\left(1-\theta\right)f_{\mathbf{x}^{\prime},n}(\lambda)
\]
and
\begin{equation}
\mathbf{f}_{\theta\mathbf{x}+(1-\theta)\mathbf{x}^{\prime},n}(\lambda)\leq\theta\mathbf{f}_{\mathbf{x},n}(\lambda)+\left(1-\theta\right)\mathbf{f}_{\mathbf{x}^{\prime},n}(\lambda).\label{eq:Footnote-Convexity}
\end{equation}
The order $\leq$ used in (\ref{eq:Footnote-Convexity}) is that on
$\Omega\subset\mathbb{R}^{m}$, inherited from the order on $\mathbb{R}^{m}$
established in Remark \ref{rmk:ControlReduction-OrderOnRN}.%
}. 
\item \label{ass:ControlReduction-ControlEnsuresMonotone}For each $\mathbf{x},\mathbf{x}^{\prime}\in\Omega$
s.t. $\mathbf{x}\leq\mathbf{x}^{\prime}$, there exist sequences $\{ \lambda_{k}\} ,\{ \lambda_{k}^{\prime}\} \in\Lambda_n^{\mathbb{N}}$
s.t. $v_{\mathbf{x},n}(\lambda_{k})\rightarrow V_{n}^{-}(\mathbf{x})$,
and for all $k$, $f_{\mathbf{x},n}(\lambda_{k})\leq f_{\mathbf{x}^{\prime},n}(\lambda_{k}^{\prime})$
and $\mathbf{f}_{\mathbf{x},n}(\lambda_{k})\leq\mathbf{f}_{\mathbf{x}^{\prime},n}(\lambda_{k}^{\prime})$.
\end{enumerate}
\begin{remark}\emph{\ref{ass:ControlReduction-ControlEnsuresMonotone}}
simplifies greatly if for all \textbf{$\mathbf{x}$}, $v_{\mathbf{x},n}(\Lambda_{n})$
contains its supremum.%
\footnote{It is worthwhile to note that in practice, this is often the case;
for fixed $n$, consider $\Lambda_{n}$ compact and $\lambda \mapsto v_{\mathbf{x},n}(\lambda)$
continuous for all $\mathbf{x}$.%
} Denote this supremum $v_{\mathbf{x},n}(\lambda_{\mathbf{x}})$,
where $\lambda_{\mathbf{x}}\in\Lambda_{n}$ is an optimal action at
$\mathbf{x}$. In this case, the following simpler assumption yields
\emph{\ref{ass:ControlReduction-ControlEnsuresMonotone}}: for each $\mathbf{x},\mathbf{x}^{\prime}\in\Omega$
s.t. $\mathbf{x}\leq\mathbf{x}^{\prime}$, there exists $\lambda^{\prime}\in\Lambda_{n}$
s.t. $f_{\mathbf{x},n}(\lambda_{\mathbf{x}})\leq f_{\mathbf{x}^{\prime},n}(\lambda^{\prime})$
and $\mathbf{f}_{\mathbf{x},n}(\lambda_{\mathbf{x}})\leq\mathbf{f}_{\mathbf{x}^{\prime},n}(\lambda^{\prime})$
(take $\lambda_{k}=\lambda_{\mathbf{x}}$ and $\lambda_{k}^{\prime}=\lambda^{\prime}$
for all $k$ to arrive at \emph{\ref{ass:ControlReduction-ControlEnsuresMonotone}}).

This simpler condition states that for each pair of positions $\mathbf{x}\leq\mathbf{x}^{\prime}$,
there is an action $\lambda^{\prime}$ s.t. the position and cash
flow after the event at $\mathbf{x}^{\prime}$ under action $\lambda^{\prime}$
are greater than (or equal to) the position and cash flow after the
event at $\mathbf{x}$ under an optimal action $\lambda_{\mathbf{x}}$.
Intuitively, this guarantees us that the position $\mathbf{x}^{\prime}$
is more desirable than $\mathbf{x}$ (from the holder's perspective).
This is not a particularly restrictive assumption, and it should hold
true for any model of a contract in which a larger position is more
desirable than a smaller one.\end{remark}
\begin{lemma}
\label{lem:ControlReduction-ConvexPreservingAcross}Suppose \emph{\ref{itm:ControlReduction-VConvexAndMonotone}},
\emph{\ref{itm:ControlReduction-BoundedFromAbove}}, and \emph{\ref{ass:ControlReduction-FlowAndStateConvexityInX}}.
Then, $\mathbf{x}\mapsto V_{n}^{-}(\mathbf{x})$ is convex.\end{lemma}
\begin{proof}
Fix $\mathbf{x},\mathbf{x}^{\prime}\in\Omega$ and $\theta\in(0,1)$,
and let $\mathbf{z}\equiv\theta\mathbf{x}+(1-\theta)\mathbf{x}^{\prime}$.
Then, by \ref{itm:ControlReduction-VConvexAndMonotone} and \ref{ass:ControlReduction-FlowAndStateConvexityInX},
\begin{align*}
V_{n}^{-}(\mathbf{z}) & =\sup v_{\mathbf{z},n}(\Lambda_{n})\\
 & =\sup_{\lambda\in\Lambda_{n}}\left[V_{n}^{+}(\mathbf{f}_{\mathbf{z},n}(\lambda))+f_{\mathbf{z},n}(\lambda)\right]\\
 & \leq\sup_{\lambda\in\Lambda_{n}}\left[V_{n}^{+}(\theta\mathbf{f}_{\mathbf{x},n}(\lambda)+\left(1-\theta\right)\mathbf{f}_{\mathbf{x}^{\prime},n}(\lambda))+\theta f_{\mathbf{x},n}(\lambda)+\left(1-\theta\right)f_{\mathbf{x}^{\prime},n}(\lambda)\right]\\
 & \leq\theta\sup_{\lambda\in\Lambda_{n}}\left[V_{n}^{+}(\mathbf{f}_{\mathbf{x},n}(\lambda))+f_{\mathbf{x},n}(\lambda)\right]+\left(1-\theta\right)\sup_{\lambda\in\Lambda_{n}}\left[V_{n}^{+}(\mathbf{f}_{\mathbf{x}^{\prime},n}(\lambda))+f_{\mathbf{x}^{\prime},n}(\lambda)\right]\\
 & =\theta\sup v_{\mathbf{x},n}(\Lambda_{n})+\left(1-\theta\right)\sup v_{\mathbf{x}^{\prime},n}(\Lambda_{n})\\
 & =\theta V_{n}^{-}(\mathbf{x})+\left(1-\theta\right)V_{n}^{-}(\mathbf{x}^{\prime}).\qedhere
\end{align*}

\end{proof}
\begin{remark}\label{rmk:ControlReduction-Statelessness}Note that
the proof of Lemma \emph{\ref{lem:ControlReduction-ConvexPreservingAcross}}
involves using $V_{n}^{-}(\mathbf{y})=\sup v_{\mathbf{y},n}(\Lambda_{n})$
for $\mathbf{y}=\mathbf{x},\mathbf{x}^{\prime}$. If $\Lambda_{n}$
is instead a function of the contract state (i.e. $\Lambda_{n}\equiv\Lambda_{n}(\mathbf{x})$),
then the above proof methodology does not work since it is not necessarily
true that $V_{n}^{-}(\mathbf{y})=\sup v_{\mathbf{y},n}(\Lambda_{n}(\mathbf{z}))$
for $\mathbf{y}=\mathbf{x},\mathbf{x}^{\prime}$.\end{remark}
\begin{lemma}
\label{lem:ControlReduction-MonotonePreservingAcross}Suppose \emph{\ref{itm:ControlReduction-VConvexAndMonotone}},
\emph{\ref{itm:ControlReduction-BoundedFromAbove}}, and \emph{\ref{ass:ControlReduction-ControlEnsuresMonotone}}.
Then, $\mathbf{x}\mapsto V_{n}^{-}(\mathbf{x})$ is monotone.\end{lemma} 
\begin{proof}
Let $\mathbf{x},\mathbf{x}^{\prime}\in\Omega$ s.t. $\mathbf{x}\leq\mathbf{x}^{\prime}$.
By \ref{itm:ControlReduction-VConvexAndMonotone} (specifically, since
$V_{n}^{+}$ is monotone) and \ref{ass:ControlReduction-ControlEnsuresMonotone},
for each $k$,
\begin{align*}
v_{\mathbf{x},n}(\lambda_{k}) & =V_{n}^{+}(\mathbf{f}_{\mathbf{x}^{\phantom{\prime}},n}(\lambda_{k}^{\phantom{\prime}}))+f_{\mathbf{x}^{\phantom{\prime}},n}(\lambda_{k}^{\phantom{\prime}})\\
 & \leq V_{n}^{+}(\mathbf{f}_{\mathbf{x}^{\prime},n}(\lambda_{k}^{\prime}))+f_{\mathbf{x}^{\prime},n}(\lambda_{k}^{\prime})\\
 & =v_{\mathbf{x}^{\prime},n}(\lambda_{k}^{\prime})
\end{align*}
Then,
\[
V_{n}^{-}(\mathbf{x})=\lim_{k\rightarrow\infty}v_{\mathbf{x},n}(\lambda_{k})\leq\limsup_{k\rightarrow\infty}v_{\mathbf{x}^{\prime},n}(\lambda_{k}^{\prime})\leq\sup v_{\mathbf{x}^{\prime},n}(\Lambda_{n})=V_{n}^{-}(\mathbf{x}^{\prime}),
\]
as desired.
\end{proof}
\begin{example}\label{exm:ControlReduction-GLWBConvexMonotoneBefore}We
now show that the GLWB satisfies \emph{\ref{ass:ControlReduction-FlowAndStateConvexityInX}}
and \emph{\ref{ass:ControlReduction-ControlEnsuresMonotone}} given \emph{\ref{itm:ControlReduction-VConvexAndMonotone}}
and \emph{\ref{itm:ControlReduction-BoundedFromAbove}}. It is trivial to
show that the functions $f_{\mathbf{x},n}^{\text{L}}(\lambda)$
and $\mathbf{f}_{\mathbf{x},n}^{\text{L}}(\lambda)$ defined
in \emph{(\ref{eq:GLWB-CashFlow})} and \emph{(\ref{eq:GLWB-StateTransition})}
are convex in $\mathbf{x}$ (the maximum of convex functions is a
convex function), thereby satisfying \emph{\ref{ass:ControlReduction-FlowAndStateConvexityInX}}.
\emph{\ref{ass:ControlReduction-ControlEnsuresMonotone}} is slightly more
tedious to verify. Let $\mathbf{x},\mathbf{x}^{\prime}\in\Omega$
s.t. $\mathbf{x}\leq\mathbf{x}^{\prime}$.  By \emph{\ref{itm:ControlReduction-VConvexAndMonotone}},
\emph{\ref{itm:ControlReduction-BoundedFromAbove}} and the argument in Example
\emph{\ref{exm:ControlReduction-GLWBBangBang}}, we can, w.l.o.g., assume
$\lambda_{\mathbf{x}}\in\{ 0,1,2\} $, where $\lambda_{\mathbf{x}}$
denotes an optimal action at $\mathbf{x}$. Hence, we need only consider
three cases:
\begin{enumerate}
\item Suppose $\lambda_{\mathbf{x}}=0$. Take $\lambda^{\prime}=0$ to get
$f_{\mathbf{x},n}^{\text{L}}(0)=f_{\mathbf{x}^{\prime},n}^{\text{L}}(\lambda^{\prime})$
and $\mathbf{f}_{\mathbf{x},n}^{\text{L}}(0)\leq\mathbf{f}_{\mathbf{x}^{\prime},n}^{\text{L}}(\lambda^{\prime})$.
\item Suppose $\lambda_{\mathbf{x}}=1$. If $x_{2}=0$, take $\lambda^{\prime}=0$.
Otherwise, $x_{2}^{\prime}\geq x_{2}>0$ and take $\lambda^{\prime}=x_{2}/x_{2}^{\prime}$.
In either case, $f_{\mathbf{x},n}^{\text{L}}(1)=f_{\mathbf{x}^{\prime},n}^{\text{L}}(\lambda^{\prime})$
and $\mathbf{f}_{\mathbf{x},n}^{\text{L}}(1)\leq\mathbf{f}_{\mathbf{x}^{\prime},n}^{\text{L}}(\lambda^{\prime})$.
\item Suppose $\lambda_{\mathbf{x}}=2$. If $x_{1}\leq\delta x_{2}$, then
$f_{\mathbf{x},n}^{\text{L}}(2)=f_{\mathbf{x},n}^{\text{L}}(1)$
and $\mathbf{f}_{\mathbf{x},n}^{\text{L}}(2)=( 0,0) \leq\mathbf{f}_{\mathbf{x},n}^{\text{L}}(1)$.
If $x_{2}=0$, take $\lambda^{\prime}=0$. Otherwise, $x_{2}^{\prime}\geq x_{2}>0$,
and once again take $\lambda^{\prime}=x_{2}/x_{2}^{\prime}$. In either case, $f_{\mathbf{x},n}^{\text{L}}(2)=f_{\mathbf{x}^{\prime},n}^{\text{L}}(\lambda^{\prime})$
and $\mathbf{f}_{\mathbf{x},n}^{\text{L}}(2)=( 0,0) \leq\mathbf{f}_{\mathbf{x}^{\prime},n}^{\text{L}}(\lambda^{\prime})$.
Therefore, we can safely assume that $x_{1}>\delta x_{2}$ so that
\begin{equation}
f_{\mathbf{x},n}^{\text{L}}(2)=\mathcal{R}(n)\left[\left(1-\kappa_{n}\right)x_{1}+\kappa_{n}\delta x_{2}\right]\leq\mathcal{R}(n)x_{1}.\label{eq:ControlReduction-BoundOnOptimalCashFlow}
\end{equation}

\begin{enumerate}
\item Suppose $x_{1}^{\prime}\leq\delta x_{2}^{\prime}$. Take $\lambda^{\prime}=1$
to get $\mathbf{f}_{\mathbf{x},n}^{\text{L}}(2)=( 0,0) \leq\mathbf{f}_{\mathbf{x}^{\prime},n}^{\text{L}}(1)$
and
\[
f_{\mathbf{x},n}^{\text{L}}(2)\leq\mathcal{R}(n)x_{1}\leq\mathcal{R}(n)\delta x_{2}^{\prime}=f_{\mathbf{x}^{\prime},n}^{\text{L}}(1)
\]
by \emph{(\ref{eq:ControlReduction-BoundOnOptimalCashFlow})}.
\item Suppose $x_{1}^{\prime}>\delta x_{2}^{\prime}$. Take $\lambda^{\prime}=2$
to get $\mathbf{f}_{\mathbf{x},n}^{\text{L}}(2)=( 0,0) =\mathbf{f}_{\mathbf{x}^{\prime},n}^{\text{L}}(2)$
and
\[
f_{\mathbf{x},n}^{\text{L}}(2)\leq\mathcal{R}(n)\left[\left(1-\kappa_{n}\right)x_{1}^{\prime}+\kappa_{n}\delta x_{2}^{\prime}\right]=f_{\mathbf{x}^{\prime},n}^{\text{L}}(2).
\]

\end{enumerate}
\end{enumerate}
\end{example}

\subsection{Preservation of convexity and monotonicity between exercise times\label{sub:ControlReduction-Between}}

As previously mentioned, to apply Theorem \ref{thm:ControlReduction-SupremumEverywhere},
we need to check the validity of \ref{itm:ControlReduction-VConvexAndMonotone}
(i.e. that the solution is CM at $t_{n}^{+}$). In
light of this, we would like to identify scenarios in which $V_{n}^{+}$
is CM provided that $V_{n+1}^{-}$ is CM
(i.e. convexity and monotonicity are preserved between exercise times).

\begin{example}\label{exm:ControlReduction-GLWBMonotoneConvexBetweenExercise}If
we assume that both GLWB and GMWB are written on an asset that follows
GBM, then Appendix \emph{\ref{app:Preservation}} establishes the convexity
and monotonicity (under sufficient regularity) of $V_{n}^{+}$ given
the convexity and monotonicity of $V_{n+1}^{-}$. The general argument
is applicable to contracts written on assets whose returns follow
multidimensional drift-diffusions with parameters independent of the
level of the asset (a local volatility model, for example, is not
included in this class). Convexity and monotonicity preservation are
retrieved directly from a property of the corresponding Green's function.

Although the methodology in Appendix \emph{\ref{app:Preservation}} relates
convexity and monotonicity to a general property of the Green's function
(including the class of contracts driven by GBM), in the interest
of intuition, we provide the reader with an alternate proof below
using the linearity of the expectation operator along with the linearity
of the stochastic process w.r.t. its initial value. Consider, in particular,
the GLWB. Equation \emph{(\ref{eq:GLWB-ValueFunction})} stipulates
\begin{align*}
V_{n}^{+}(\mathbf{x}) & =\tilde{\mathbb{E}}\Bigl[e^{-\int_{n}^{n+1}r(\tau)d\tau}V_{n+1}^{-}(X_{1}(\left(n+1\right)^{-}),x_{2})\\
 & \qquad\qquad+\int_{n}^{n+1}e^{-\int_{n}^{s}r(\tau)d\tau}\mathcal{M}(s)X_{1}(s)ds\mid X_{1}(n^{+})=x_{1}\Bigr] & \text{ on }\left[0,\infty\right)^{2}\times\mathscr{T}.
\end{align*}
Linearity allows us to consider the two terms appearing in the sum
inside the conditional expectation separately. If each is convex in
$\mathbf{x}$, so too is the entire expression. If $X_{1}(n^{+})=x_{1}$,
\[
X_{1}(s)=x_{1}Y(s)
\]
between $n$ and $n+1$, where
\[
Y(s)\equiv\exp\biggl(\int_{n}^{s}\left[r(\tau)-\alpha(\tau)-\frac{1}{2}\sigma^{2}(\tau)\right]d\tau+\int_{n}^{s}\sigma(\tau)d\tilde{Z}(\tau)\biggr),
\]
from which it is evident that $X_{1}(s)$ is convex 
in $x_{1}$ since $Y$ depends only on time (note that the
parameters appearing in $Y$ are independent of the level of the asset,
precluding a local volatility model). It remains to show that the
first term is also convex. Fix $\mathbf{y},\mathbf{y}^{\prime}\in[0,\infty)^{2}$,
$\theta\in(0,1)$, and let $\mathbf{x}\equiv\theta\mathbf{y}+(1-\theta)\mathbf{y}^{\prime}$.
Then, by assuming that $V_{n+1}^{-}(\mathbf{x})$ is convex
in $\mathbf{x}$,
\begin{align*}
V_{n+1}^{-}(x_{1}Y(\left(n+1\right)^{-}),x_{2}) & =V_{n+1}^{-}((\theta y_{1}+\left(1-\theta\right)y_{1}^{\prime})Y(\left(n+1\right)^{-}),\theta y_{2}+\left(1-\theta\right)y_{2}^{\prime})\\
 & \leq\theta V_{n+1}^{-}(y_{1}Y(\left(n+1\right)^{-}),y_{2})+\left(1-\theta\right)V_{n+1}^{-}(y_{1}^{\prime}Y(\left(n+1\right)^{-}),y_{2}^{\prime}).
\end{align*}
One can use the same technique to show that monotonicity is preserved.
An identical argument can be carried out for the GMWB.\end{example}

Convexity and monotonicity preservation are established for a stochastic
volatility model in \cite{bergman1996general}. For the case of general
parabolic equations, convexity preservation is established in \cite{janson2004preservation}.
This result is further generalized to parabolic integro-differential
equations, arising from problems involving assets whose returns follow
jump-diffusion processes \cite{bian2008convexity}.

\subsection{Existence of an optimal bang-bang control}

Once we have established that convexity and monotonicity are preserved
across and between exercise times (i.e. $\S$\ref{sub:ControlReduction-PreservationAcross}
and $\S$\ref{sub:ControlReduction-Between}, respectively), we need
only apply our argument inductively to establish the existence of
an optimal bang-bang control. Instead of providing a proof for the
general case, we simply focus on the GLWB contract here. For the case
of a general contract, assuming the dynamics followed by the assets
preserve the convexity and monotonicity of the cost of funding the
contract between exercise times (e.g. GBM, as in Appendix \ref{app:Preservation}),
the reader can apply the same techniques to establish the existence
of a bang-bang control.

\begin{example}Consider the GLWB. Suppose that for some $n$ s.t.
$0\leq n<N$, $V_{n+1}^{-}$ is CM. By Example
\emph{\ref{exm:ControlReduction-GLWBMonotoneConvexBetweenExercise}},
$V_{n}^{+}$ is also CM. Under sufficient regularity
(see Appendix \emph{\ref{app:Preservation}}), for fixed $\mathbf{x}$, $v_{\mathbf{x},n}$
is bounded above (satisfying \emph{\ref{itm:ControlReduction-BoundedFromAbove}}).
Since \emph{\ref{itm:ControlReduction-VConvexAndMonotone}} and \emph{\ref{itm:ControlReduction-BoundedFromAbove}}
are satisfied, we can use Example \emph{\ref{exm:ControlReduction-GLWBBangBang}}
to conclude that the supremum of $v_{\mathbf{x},n}$, for each $\mathbf{x}\in\Omega$,
occurs on $\{ 0,1,2\} $. By Example \emph{\ref{exm:ControlReduction-GLWBConvexMonotoneBefore}},
$V_{n}^{-}$ is convex and monotone.

By \emph{(\ref{eq:GLWB-Initial})} and \emph{(\ref{eq:GLWB-Initial2})},
$V(\mathbf{x},N)=0$.
Since $V(\mathbf{x},N)$ is trivially CM
as a function of $\mathbf{x}$, we can apply the above argument inductively
to establish the existence of an optimal bang-bang control. \end{example}

\section{Numerical Examples}\label{sec:NumericalResults}

To demonstrate the bang-bang principle in practice, we implement a
numerical method to solve the GLWB and GMWB problems and examine loss-maximizing
withdrawal strategies.

\subsection{Contract pricing algorithm}

Algorithm \ref{alg:NumericalMethod-Algorithm} highlights the usual
dynamic programming approach to pricing contracts with finitely many
exercise times. Note that line 2 
is purposely non-specific;
the algorithm does not presume anything about the underlying dynamics
of the stochastic process(es) that $V$ is a function of, and as such
does not make mention of a particular numerical method used to solve
$V_{n}^{+}$ given $V_{n+1}^{-}$. Establishing that the control is bang-bang
for a particular contract allows us to replace $\Lambda_{n}$ appearing
on line 4 
with a finite subset of itself.

\begin{algorithm}
 \caption{Dynamic programming for pricing contracts with finitely many exercise times.\label{alg:NumericalMethod-Algorithm}}
 \vskip\belowcaptionskip
 \LinesNumbered
 \KwData{payoff at the expiry, $V_N=\varphi$}
 \KwResult{price of the contract at time zero, $V_0\equiv V_0^-$}
 \For{$n\leftarrow N - 1$ \KwTo $0$}{
  use $V_{n+1}^-$ to determine $V_n^+$ \label{alg:Between} \\
  \For{$\mathbf{x} \in \Omega$}{
    $
    V_n^-(\mathbf{x})
    \equiv
    \sup_{\lambda\in\Lambda_{n}}V_{n}^+(\mathbf{f}_{\mathbf{x},n}(\lambda))+f_{\mathbf{x},n}(\lambda)
   $ \label{alg:Across}
  }
 }
\end{algorithm}

\subsection{Numerical method}

The numerical method discussed here applies to both GLWB and GMWB
contracts. Each contract is originally posed on $\Omega=[0,\infty)^{2}$.
We employ Algorithm \ref{alg:NumericalMethod-Algorithm} but instead
approximate the solution using a finite difference method on the truncated
domain $[0,x_{1}^{\text{max}}]\times[0,x_{2}^{\text{max}}]$.
As such, since $\mathbf{f}_{\mathbf{x},n}(\lambda)$ will
not necessarily land on a mesh node, linear interpolation is used
to approximate $V_{n}^{+}(\mathbf{f}_{\mathbf{x},n}(\lambda))$
on line 4.  
A local optimization problem is solved
for each point on the finite difference grid. Details of the numerical
scheme can be found in \cite{azimzadeh2013hedging,forsyth2013risk}.

Between exercise times, the cost of funding each contract satisfies
one of (\ref{eq:GMWB-PDE}) or (\ref{eq:GLWB-PDE}). Corresponding
to line 2 
of Algorithm \ref{alg:NumericalMethod-Algorithm},
we determine $V_{n}^{+}$ from $V_{n+1}^{-}$ using an implicit finite
difference discretization. No additional boundary condition is needed
at $x_{1}=0$ or $x_{2}=0$ ((\ref{eq:GMWB-PDE}) and (\ref{eq:GLWB-PDE})
hold along $\partial\Omega\times[t_{n},t_{n+1})$). The
same is true of $x_{2}=x_{2}^{\text{max}}\gg0$.  At $x_{1}=x_{1}^{\text{max}}\gg0$,
we impose
\begin{equation}
V(x_{1}^{\text{max}},x_{2},t)=g(t)x_{1}^{\text{max}}\label{eq:NumericalResults-Asymptotic}
\end{equation}
for some function $g$ differentiable everywhere but possibly
at the exercise times $n$. Substituting the above into (\ref{eq:GMWB-PDE})
or (\ref{eq:GLWB-PDE}) yields an ordinary differential equation which
is solved numerically alongside the rest of the domain. Errors introduced
by the above approximations are small in the region of interest, as
verified by numerical experiments.

\begin{remark}Since we advance the numerical solution from $n^{-}$
to $(n-1)^{+}$ using a convergent method, the numerical
solution converges pointwise to a solution $V$ that is convexity
and monotonicity preserving. Although it is possible to show---for
special cases---that convexity and monotonicity are preserved for
finite mesh sizes, this is not necessarily true unconditionally.\end{remark}

\begin{remark}Although we have shown that an optimal bang-bang control
exists for the GLWB problem, we do not replace $\Lambda_{n}$ with
$\{ 0,1,2\} $ on line \emph{4} 
of Algorithm \emph{\ref{alg:NumericalMethod-Algorithm}} when computing the
cost to fund a GLWB in $\S$\emph{\ref{sub:Results-GLWB}} so as to demonstrate
that our numerical method, having preserved convexity and monotonicity,
selects an optimal bang-bang control. For both GLWB and GMWB, We assume
that nothing is known about $v_{\mathbf{x},n}$ and hence form a partition
\[
\lambda_{1}<\lambda_{2}<\cdots<\lambda_{p}
\]
of the admissible set and perform a linear search%
\end{remark}

\subsection{Results}

\subsubsection{Guaranteed Lifelong Withdrawal Benefits\label{sub:Results-GLWB}}

\begin{figure}
\caption{Optimal control for the GLWB for data in Table \emph{\ref{tab:Results-GLWBParameters}}.
As predicted, there exists an optimal control consisting only of nonwithdrawal,
withdrawal at the contract rate, and a full surrender.\label{fig:Results-GLWBWithdrawalStrategies}}
\vskip\belowcaptionskip
\centering

\includegraphics[height=0.175in]{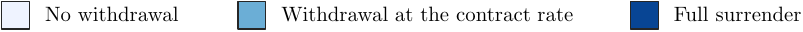}

\subfigure[$n=1$]{

\includegraphics[width=2.9in]{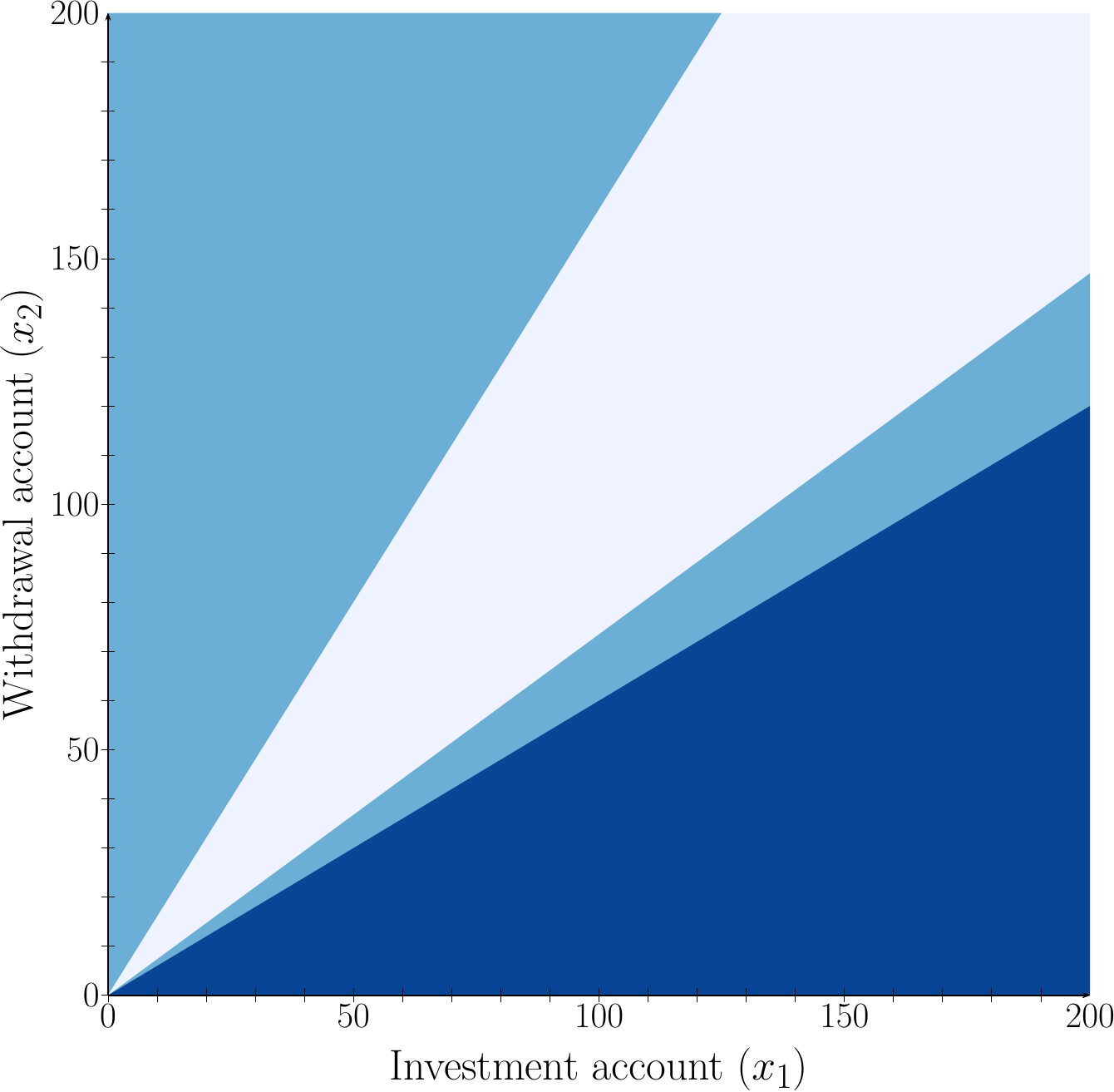}

}
\subfigure[$n=2$]{

\includegraphics[width=2.9in]{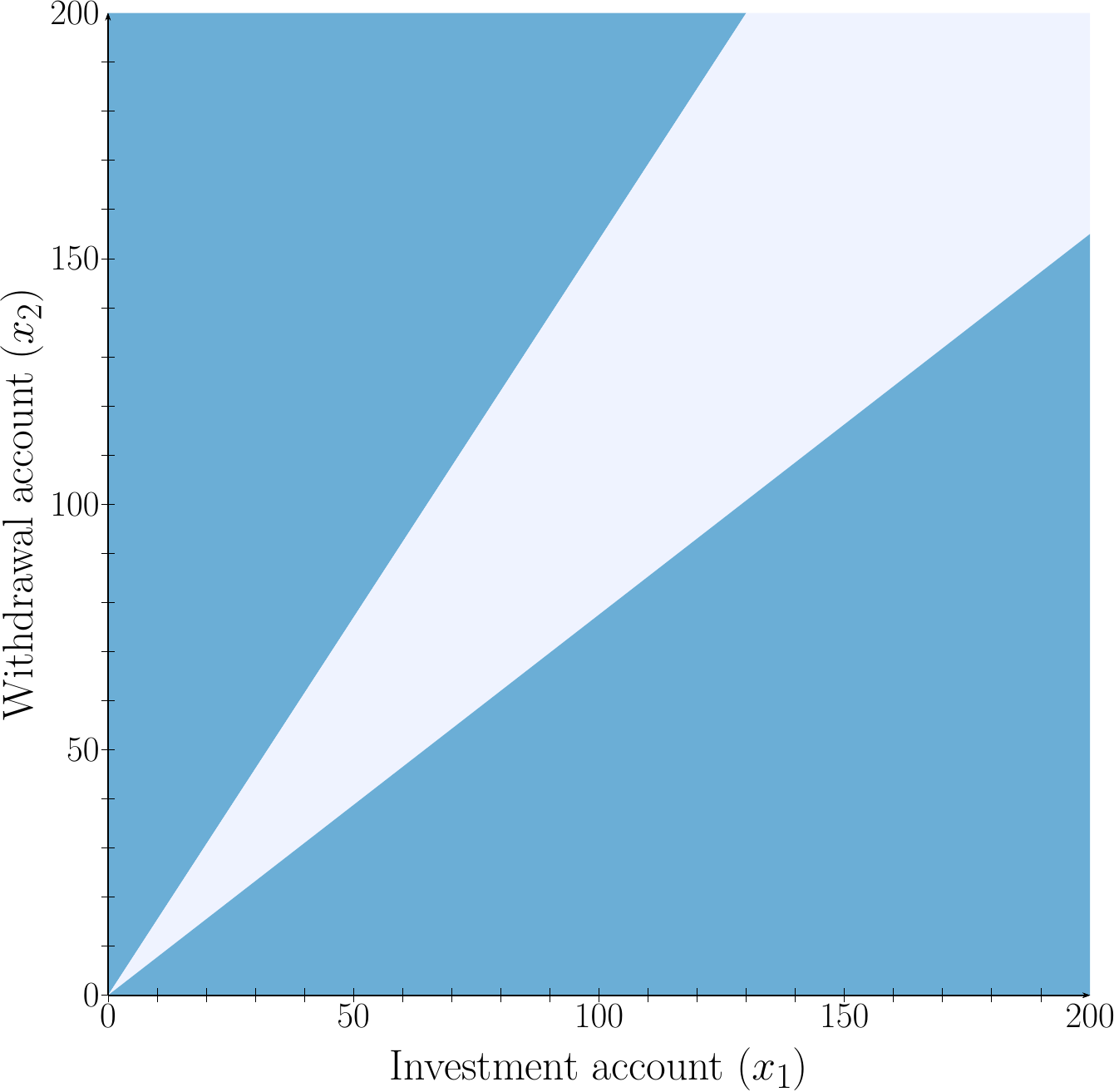}

}
\par\medskip
\subfigure[$n=3$]{

\includegraphics[width=2.9in]{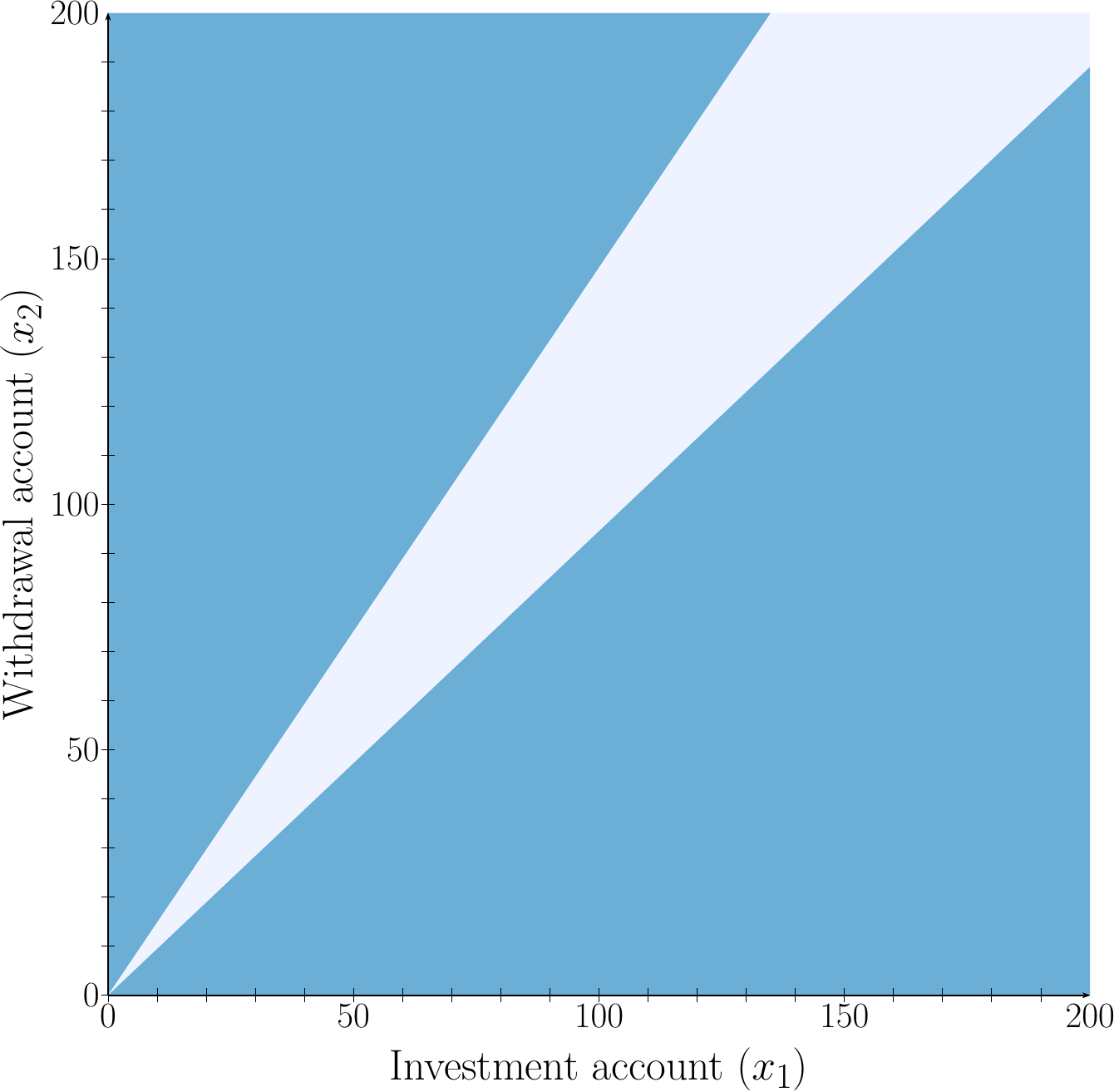}

}
\subfigure[$n=4$]{

\includegraphics[width=2.9in]{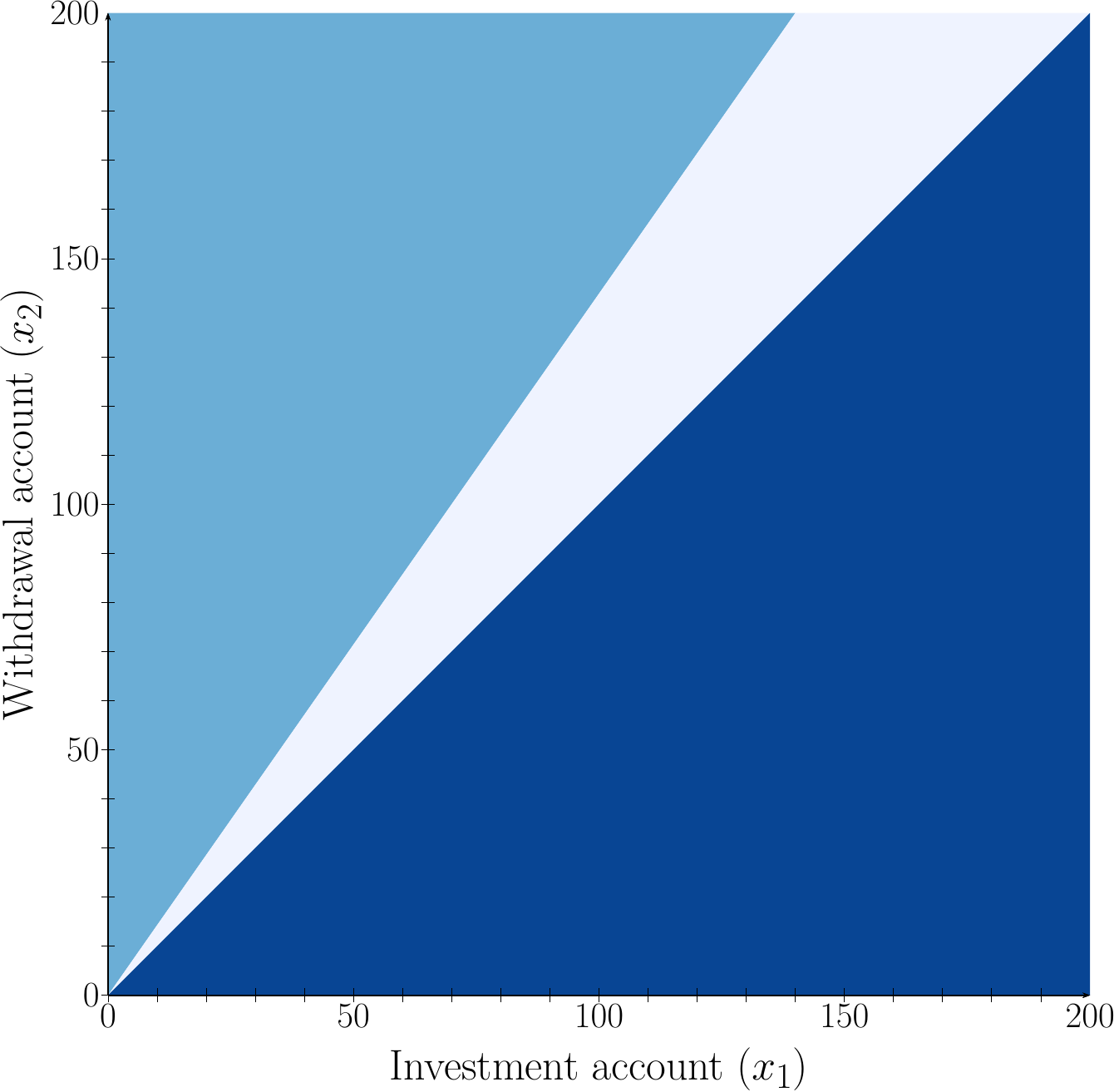}

}
\end{figure}

Figure \ref{fig:Results-GLWBWithdrawalStrategies} shows withdrawal
strategies for the holder under the parameters in Table \ref{tab:Results-GLWBParameters}
on the first four contract anniversaries. We can clearly see that
the optimal control is bang-bang from the figures. At any point $(\mathbf{x},n)$,
we see that the holder performs one of nonwithdrawal, withdrawal at
exactly the contract rate, or a full surrender (despite being afforded
the opportunity to withdraw any amount between nonwithdrawal and a full
surrender).

When the withdrawal benefit is much larger than the investment account,
the optimal strategy is withdrawal at the contract rate (the guarantee
is in the money). Conversely, when the investment account is much
larger than the withdrawal benefit, the optimal strategy is surrender
(the guarantee is out of the money), save for when the holder is anticipating
the triennial ratchet (times $n=2$ and $n=3$). Otherwise, the optimal
strategy includes nonwithdrawal (to receive a bonus) or withdrawal
at the contract rate. Note that the strategy is constant along any
straight line through the origin since the solution is homogeneous
of order one in $\mathbf{x}$, as discussed in \cite{forsyth2013risk}.
\begin{table}
\protect\caption{GLWB parameters.\label{tab:Results-GLWBParameters}}

\centering{}%
\begin{tabular}{lcr}
\toprule
\multicolumn{2}{l}{\textbf{Parameter}} & \textbf{Value}\tabularnewline
\midrule
Volatility & $\sigma$ & 0.20\tabularnewline
\midrule
Risk-free rate & $r$ & 0.04\tabularnewline
\midrule
Hedging fee & $\alpha$ & 0.015\tabularnewline
\midrule
Contract rate & $\delta$ & 0.05\tabularnewline
\midrule
Bonus rate & $\beta$ & 0.06\tabularnewline
\midrule
Expiry & $N$ & 57\tabularnewline
\midrule
Initial investment & $w_{0}$ & 100\tabularnewline
\midrule
Initial age at time zero &  & 65\tabularnewline
\midrule
Mortality data &  & \cite{pasdika2005coping}\tabularnewline
\midrule
Ratchets &  & Triennial\tabularnewline
\midrule
Withdrawals &  & Annual\tabularnewline
\bottomrule
\end{tabular}
\quad{}
\begin{tabular}{lr}
\toprule
\textbf{Anniversary $n$} & \textbf{Penalty $\kappa_{n}$}\tabularnewline
\midrule
1 & 0.03\tabularnewline
\midrule
2 & 0.02\tabularnewline
\midrule
3 & 0.01\tabularnewline
\midrule
$\geq4$ & 0.00\tabularnewline
\bottomrule
\end{tabular}
\end{table}

\subsubsection{\label{sub:Results-GMWB}Guaranteed minimum withdrawal benefit}

For the GMWB, \ref{ass:ControlReduction-FlowAndStateConvexityInX}
is violated. In particular, for $\kappa_{n}>0$, the function $f_{\mathbf{x},n}^{\text{M}}(\lambda)$
is concave as a function of $\mathbf{x}$. However, when $\kappa_{n}=0$
or $G=0$ ($G=0$ is considered in \cite{huang2013analysis}), the
function $f_{\mathbf{x},n}^{\text{M }}(\lambda)$ (see
(\ref{eq:GMWB-CashFlow})) is linear in $\mathbf{x}$, and hence the
convexity of $V_{n}^{-}$ can be guaranteed given $V_{n}^{+}$ CM.
In this case, it is possible to use the same machinery
as was used in the GLWB case to arrive at a bang-bang principle (see
Theorem \ref{thm:ControlReduction-SupremumEverywhere}). The
case of $\kappa_{n}=0$ corresponds to zero surrender charges at the
$n$th anniversary, while $G=0$ corresponds to enforcing
that all withdrawals (regardless of size) be charged at the penalty
rate.

\begin{figure}
\caption{$V(\mathbf{x},t)$ for fixed $x_{1}=100$ under the data
in Table \emph{\ref{tab:Results-GMWBParameters}}. Points where $V(\mathbf{x},n^{-})=V(\mathbf{x},n^{+})$
correspond to nonwithdrawal. To the left of these points, the holder
performs withdrawal (see Figure \emph{\ref{fig:Results-GMWBWithdrawalStrategies}}).\label{fig:Results-GMWBConvexity}}
\centering
\subfigure[Convexity is not preserved from $t\rightarrow6^{+}$ to $t\rightarrow6^{-}$.]{

\includegraphics[width=2.9in]{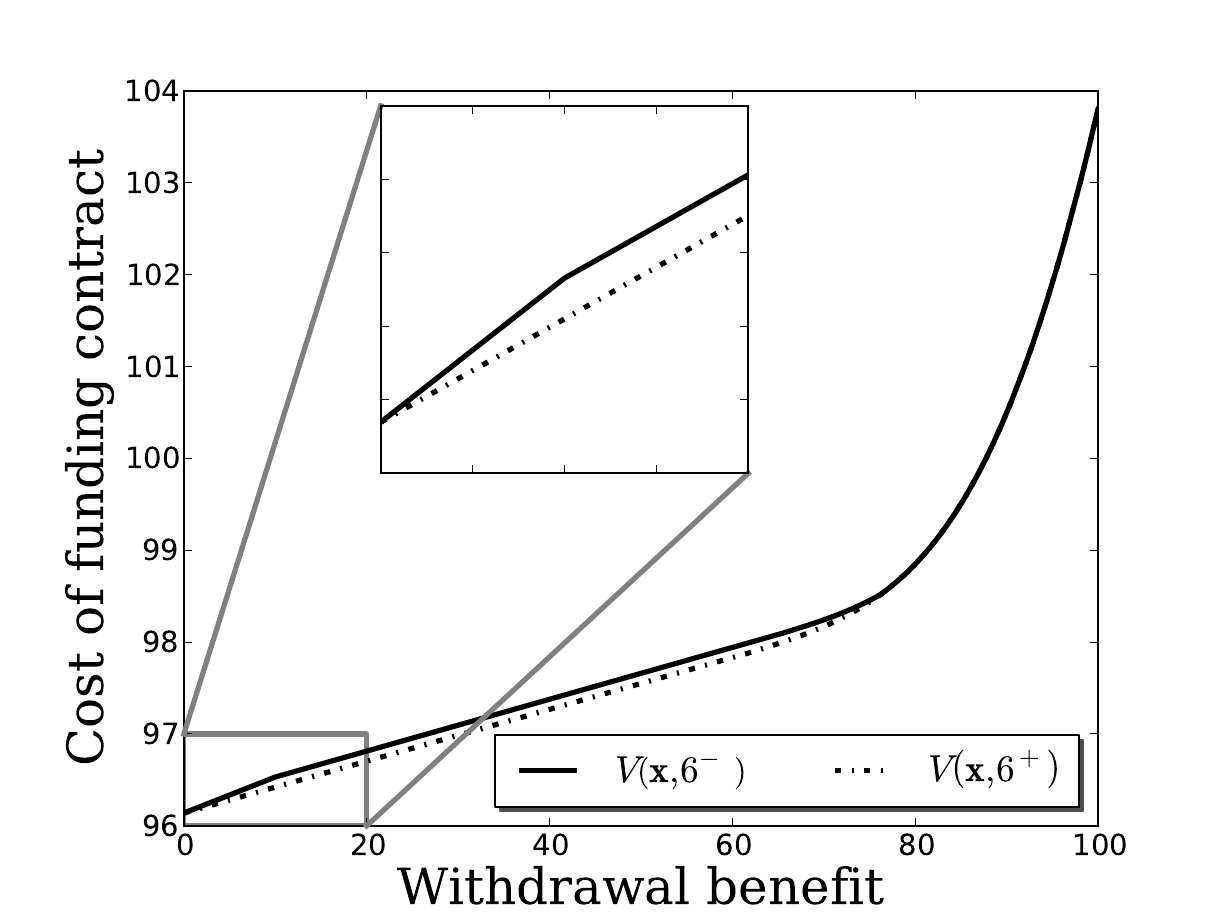}

}
\subfigure[Convexity is preserved from $t\rightarrow7^{+}$ to $t\rightarrow7^{-}$.]{

\includegraphics[width=2.9in]{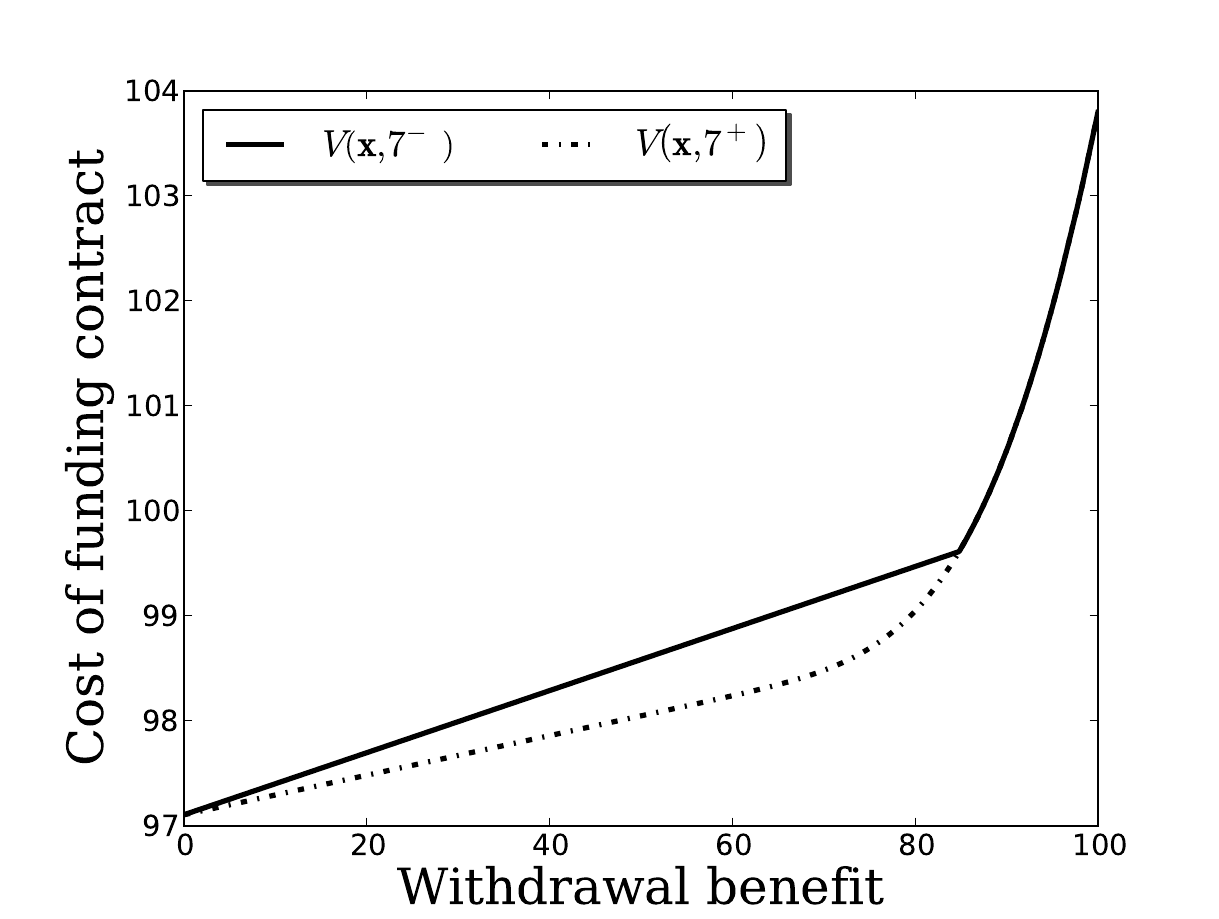}}
\end{figure}

Now, consider the data in Table \ref{tab:Results-GMWBParameters}.
Since $\kappa_{n}=0$ for all $n\geq7$, the convexity of $V$ in
$\mathbf{x}$ is preserved for all $t\in(6,N]$. However,
since $\kappa_{6}>0$, the convexity is violated as $t\rightarrow6^{-}$.
Figure \ref{fig:Results-GMWBConvexity} demonstrates this preservation
and violation of convexity. As a consequence, $V$ will not necessarily
be convex in $\mathbf{x}$ as $t\rightarrow5^{+}$, and the contract
fails to satisfy the bang-bang principle at each anniversary date
$n\le5$.

\begin{figure}
\caption{Optimal control $\lambda_{\mathbf{x}}$ scaled by $x_{2}$ for the
data in Table \emph{\ref{tab:Results-GMWBParameters}}.\label{fig:Results-GMWBWithdrawalStrategies}}
\vskip\belowcaptionskip

\begin{center}
\includegraphics[height=0.175in]{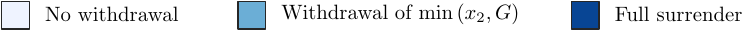}
\par\end{center}

\centering
\subfigure[$n=6$]{

\includegraphics[width=2.9in]{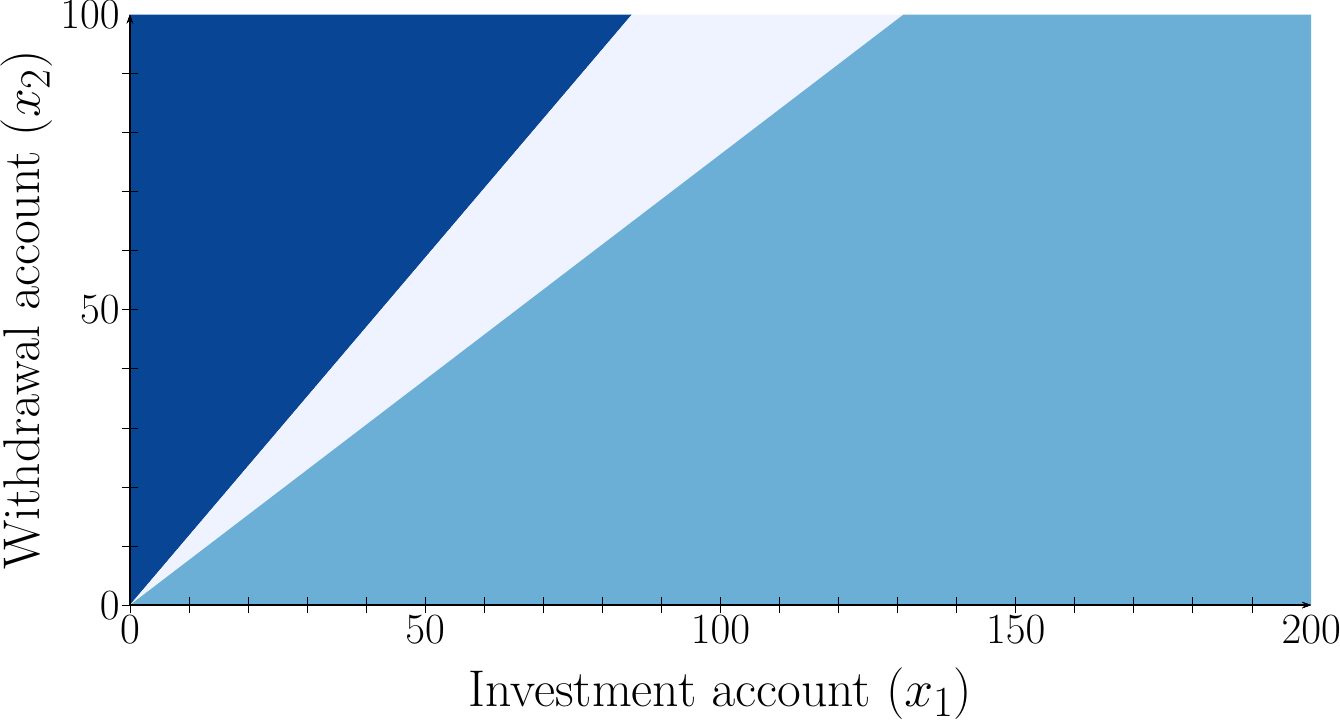}

}
\subfigure[$n=7$]{

\includegraphics[width=2.9in]{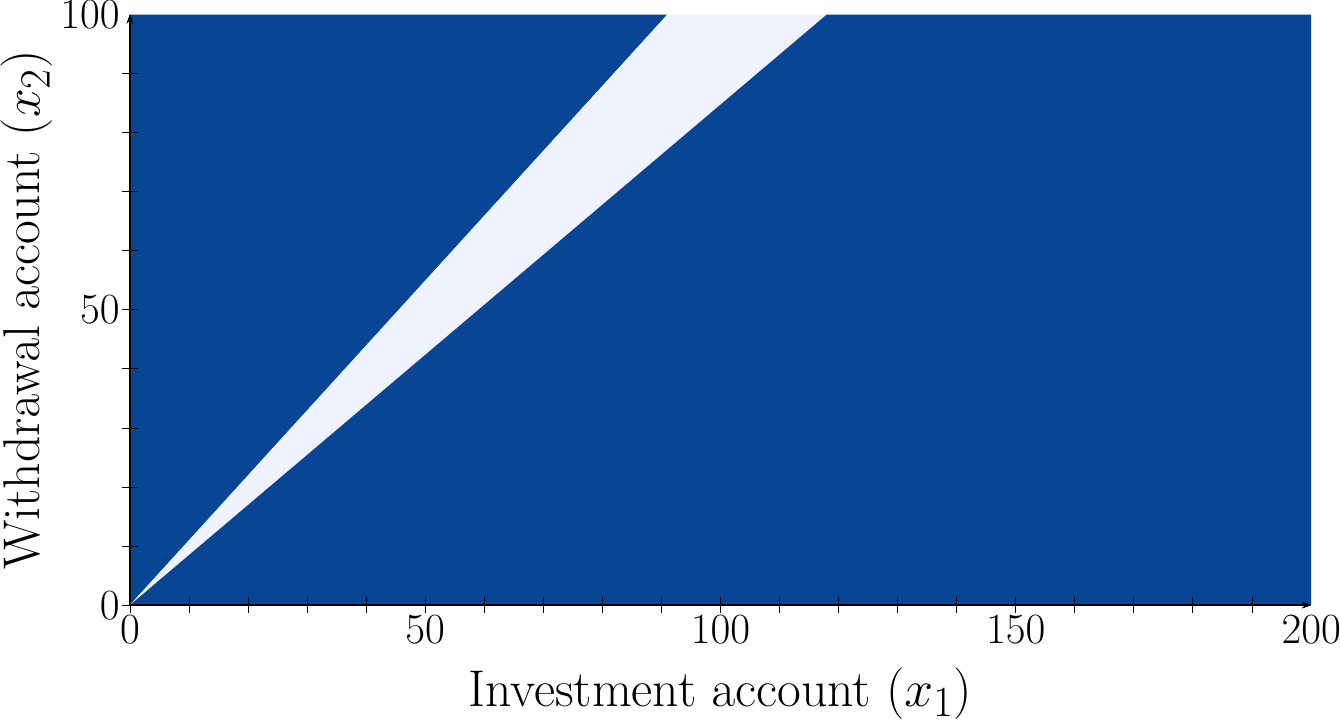}

}
\end{figure}

Note that for $x_{2}>0$, the conditions $\lambda x_{2}\in[0,G\wedge x_{2}]$
and $\lambda x_{2}\in(G\wedge x_{2},x_{2}]$ appearing
in (\ref{eq:GMWB-CashFlow}) are equivalent to $\lambda\in[0,G/x_{2}\wedge1]$
and $\lambda\in(G/x_{2}\wedge1,1]$, respectively. Assuming
that $V_{n}^{+}$ is CM  and taking $\mathcal{P}_{n}^{\text{M}}(\mathbf{x})\equiv\{ P_{1},P_{2}\} $
with $P_{1}\equiv[0,G/x_{2}\wedge1]$ and $P_{2}\equiv[G/x_{2}\wedge1,1]$
yields that there exists an optimal control taking on one of the values
in $\{ 0,G/x_{2},1\} $ at any point $(\mathbf{x},n)$
with $x_{2}>0$. These three actions correspond to nonwithdrawal,
withdrawing the predetermined amount $G$, or performing a full surrender.
This is verified by Figure \ref{fig:Results-GMWBWithdrawalStrategies},
which shows withdrawal strategies under the parameters in Table \ref{tab:Results-GMWBParameters}
at times $n=6$ and $n=7$. As predicted, along any line $x_{2}=\text{const.}$,
the optimal control takes on one of a finite number of values. Since
at $n=6$, $\kappa_{n}>0$, we see that the holder is more hesitant
to surrender the contract whenever $x_{1}\gg x_{2}$ (compare with
the same region at $n=7$). Control figures for GMWBs not satisfying
the bang-bang principle can be seen in the numerical results in \cite{dai2008guaranteed,chen08a}.

\begin{table}
\protect\caption{GMWB parameters \emph{\cite{chen2008effect}}.\label{tab:Results-GMWBParameters}}

\centering{}%
\begin{tabular}{lcr}
\toprule
\multicolumn{2}{l}{\textbf{Parameter}} & \textbf{Value}\tabularnewline
\midrule
Volatility & $\sigma$ & 0.15\tabularnewline
\midrule
Risk-free rate & $r$ & 0.05\tabularnewline
\midrule
Hedging fee & $\alpha$ & 0.01\tabularnewline
\midrule
Contract rate & $G$ & 10\tabularnewline
\midrule
Expiry & $N$ & 10\tabularnewline
\midrule
Initial investment & $w_{0}$ & 100\tabularnewline
\midrule
Withdrawals &  & Annual\tabularnewline
\bottomrule
\end{tabular}
\quad{}%
\begin{tabular}{lr}
\toprule
\textbf{Anniversary $n$} & \textbf{Penalty $\kappa_{n}$}\tabularnewline
\midrule
1 & 0.08\tabularnewline
\midrule
2 & 0.07\tabularnewline
\midrule
3 & 0.06\tabularnewline
\midrule
4 & 0.05\tabularnewline
\midrule
5 & 0.04\tabularnewline
\midrule
6 & 0.03\tabularnewline
\midrule
$\geq7$ & 0.00\tabularnewline
\bottomrule
\end{tabular}
\end{table}

\section{Conclusion}

Although it is commonplace in the insurance literature to assume the
existence of optimal bang-bang controls, there does not appear to
be a rigorous statement of this result. We have rigorously derived
sufficient conditions which guarantee the existence of optimal bang-bang
controls for GMxB guarantees.

These conditions require that the contract features be such that
the solution to the optimal control can be formulated as maximizing
a convex objective function, and that the underlying stochastic process
assumed for the risky assets preserves convexity and monotonicity.

These conditions are non-trivial, in that the conditions are satisfied
for the GLWB contract but not for the GMWB contract with typical contract
parameters.
From a practical point of view, the existence of optimal
bang-bang controls allows for
the use of very efficient numerical methods.

Although we have focused specifically on the application of our results
to GMxB guarantees, the reader will have no difficulty in applying
the sufficient conditions to other optimal control problems in finance.
We believe that we can also use an approach similar to that used here
to establish the existence of optimal bang-bang controls
for general impulse control problems.
In the impulse control case, these conditions will require that the
intervention operator have a particular form and that the stochastic
process (without intervention) preserve convexity and monotonicity.
We leave this generalization for future work.

\appendix

\section{Preservation of convexity and monotonicity\label{app:Preservation}}

In this appendix, we establish the convexity and monotonicity of a
contract whose payoff is CM and written on assets
whose returns follow (multidimensional, possibly correlated) GBM.
We do so by considering the PDE satisfied by $V$ and the fundamental
solution corresponding to the operator appearing in the log-transformed
version of this PDE. Considering the log-transformed PDE allows us
to eliminate the parabolic degeneracy at the boundaries and to argue
that the resulting fundamental solution for the log-transformed operator
should be of the form $\Gamma(\mathbf{y},\mathbf{y}^{\prime},\tau,\tau^{\prime})\equiv\Gamma(\mathbf{y}-\mathbf{y}^{\prime},\tau,\tau^{\prime})$.

We begin by describing some of the notation used in this appendix:
\begin{itemize}
\item Let $\Omega\equiv\Omega_{1}\times\Omega_{2}$ where $\Omega_{1}\equiv(0,\infty)^{m}$
and $\Omega_{2}$ is a convex subset of a partially ordered vector
space $A$ over  $\mathbb{R}$ with order $\leq_{A}$. $\Omega$
can thus be considered as a convex subset of the vector space $\mathcal{A}\equiv\mathbb{R}^{m}\times A$
over  $\mathbb{R}$.
\item We write an element of $\Omega$ in the form $(\mathbf{x},x_{m+1})\equiv(x_{1},\ldots,x_{m},x_{m+1})$
with $\mathbf{x}\in\Omega_{1}$ and $x_{m+1}\in\Omega_{2}$ in order
to distinguish between elements of $\Omega_{1}$ and $\Omega_{2}$.
\item The partial order we consider on $\mathcal{A}$ is simply inherited
from the orders $\leq$ on $\mathbb{R}^{m}$ (Remark \ref{rmk:ControlReduction-OrderOnRN})
and $\leq_{A}$. Specifically, $(\mathbf{x},x_{m+1})\leq_{\mathcal{A}}(\mathbf{x}^{\prime},x_{m+1}^{\prime})$
if and only if $\mathbf{x}\leq_{\mathbb{R}^{m}}\mathbf{x}^{\prime}$
and $x_{m+1}\leq_{A}x_{m+1}^{\prime}$.
\end{itemize}
Suppose $V$ satisfies

\begin{equation}
\partial_{t}V+\mathcal{L}V+\omega=0\text{ on }\Omega\times\left(t_{n},t_{n+1}\right)\label{eq:Preservation-PDE}
\end{equation}
and
\begin{equation}
V(\mathbf{x},x_{m+1},t_{n+1}^{-})=\varphi(\mathbf{x},x_{m+1})\text{ on }\Omega\label{eq:Preservation-Cauchy}
\end{equation}
where
\begin{equation}
\mathcal{L}\equiv\frac{1}{2}\sum_{i,j=1}^{m}a_{i,j}x_{i}x_{j}\partial_{x_{i}}\partial_{x_{j}}+\sum_{i=1}^{m}b_{i}x_{i}\partial_{x_{i}}+c.\label{eq:Preservation-Operator}
\end{equation}
In the above, $\omega\equiv\omega(\mathbf{x},x_{m+1},t)$. We
will, for the remainder of this appendix, assume the following:
\begin{enumerate}[label=(D\arabic*)]
\item \label{itm:Preservation-Independence}$a_{i,j}\equiv a_{i,j}(t)$,
$b_{i}\equiv b_{i}(t)$, and $c\equiv c(t)$
(i.e. the functions $a_{i,j}$, $b_{i}$ and $c$ are independent
of $(\mathbf{x},x_{m+1})$).
\item \label{itm:Preservation-UniformlyParabolic}
$\sum_{i,j=1}^m a_{i,j} \partial_{x_i} \partial_{x_j}$ is uniformly elliptic.
\end{enumerate}
\begin{example}For the GLWB guarantee, $\mathcal{L}$ is given in
\emph{(\ref{eq:GMWB-L})} and $\omega=\mathcal{M}(t)x_{1}$.\end{example}

\begin{remark}We say $V$ satisfies \emph{(\ref{eq:Preservation-PDE})}
if $V$ is twice differentiable in (the components of) $\mathbf{x}$
and once differentiable in $t$ on $\Omega\times(t_{n},t_{n+1})$,%
\footnote{i.e. $V|_{\Omega\times(t_{n},t_{n+1})}\in C^{2,1}(\Omega\times(t_{n},t_{n+1}))$.%
} continuous on $\Omega\times(t_{n},t_{n+1}]$,%
\footnote{i.e. $V|_{\Omega\times(t_{n},t_{n+1}]}\in C(\Omega\times(t_{n},t_{n+1}])$.%
} and satisfies \emph{(\ref{eq:Preservation-PDE})} pointwise.\end{remark}

We now describe the log-transformed problem. For ease of notation,
let
\begin{align*}
e^{\mathbf{y}} & \equiv\left(e^{y_{1}},\ldots,e^{y_{m}}\right) & a_{i,j}^{\prime}(\tau) & \equiv a_{i,j}(t_{n+1}-\tau)\\
\varphi^{\prime}(\mathbf{y},y_{m+1}) & \equiv\varphi(e^{\mathbf{y}},y_{m+1}) & b_{i\phantom{,j}}^{\prime}(\tau) & \equiv b_{i}(t_{n+1}-\tau)\\
\omega^{\prime}(\mathbf{y},y_{m+1},\tau) & \equiv\omega(e^{\mathbf{y}},y_{m+1},t_{n+1}-\tau) & c_{\phantom{i,j}}^{\prime}(\tau) & \equiv c(t_{n+1}-\tau)
\end{align*}
and $\Omega^{\prime}\equiv\mathbb{R}^{m}\times\Omega_{2}$. Let $V$
be a solution of the Cauchy problem (\ref{eq:Preservation-PDE}) and
(\ref{eq:Preservation-Cauchy}). Let
\[
u(\mathbf{y},y_{m+1},\tau)\equiv V(e^{\mathbf{y}},y_{m+1},t_{n+1}-\tau)
\]
 and $\Delta\equiv t_{n+1}-t_{n}$. Then, $u$ satisfies
\begin{equation}
\mathcal{L}^{\prime}u-\partial_{\tau}u+\omega^{\prime}=0\text{ on }\Omega^{\prime}\times\left(0,\Delta\right)\label{eq:Preservation-LogPDE}
\end{equation}
and
\begin{equation}
u(\mathbf{y},y_{m+1},0)=\varphi^{\prime}(\mathbf{y},y_{m+1})\label{eq:Preservation-LogCauchy}
\end{equation}
where
\[
\mathcal{L}^{\prime}\equiv\frac{1}{2}\sum_{i,j=1}^{m}a_{i,j}^{\prime}\partial_{y_{i}}\partial_{y_{j}}+\sum_{i=1}^{m}\left(b_{i}^{\prime}-\frac{1}{2}a_{i,i}^{\prime}\right)\partial_{y_{i}}+c^{\prime}.
\]
Note that \ref{itm:Preservation-UniformlyParabolic} implies that
$\mathcal{L}^{\prime}$ is uniformly elliptic. 

In order to guarantee that a solution $u$ to the log-transformed
Cauchy problem (\ref{eq:Preservation-LogPDE}) and (\ref{eq:Preservation-LogCauchy})
exists, and is unique, sufficient regularity must be imposed on
$\varphi^{\prime}$, $\mathcal{L}^{\prime}$, and $\omega^{\prime}$.
We summarize below.
\begin{enumerate}[label=(E\arabic*)]
\item \label{itm:Preservation-Continuous}For each $y_{m+1}$, $\mathbf{y}\mapsto\varphi^{\prime}(\mathbf{y},y_{m+1})$
is continuous on $\mathbb{R}^{m}$.
\item \label{itm:Preservation-Regularity}The coefficients of $\mathcal{L}^{\prime}$
are sufficiently regular.
\item \label{itm:Preservation-wRegular}For each $y_{m+1}\in\Omega_{2}$,
$(\mathbf{y},\tau)\mapsto\omega^{\prime}(\mathbf{y},y_{m+1},\tau)$ is sufficiently
regular.
\item \label{itm:Preservation-Growth} $u$ satisfies a growth condition as
  $|\mathbf{y}| \rightarrow \infty$.
\end{enumerate}
For an accurate detailing of the required regularity, see \cite[Chap. 1: Thms. 12 and 16]{friedmanpartial}.

When \ref{itm:Preservation-UniformlyParabolic} and \ref{itm:Preservation-Continuous}---\ref{itm:Preservation-Growth}
are satisfied, the solution $u$ can be written as
\begin{multline}
u(\mathbf{y},y_{m+1},\tau)=\int_{\mathbb{R}^{m}}\Gamma(\mathbf{y},\mathbf{y}^{\prime},\tau,0)\varphi^{\prime}(\mathbf{y}^{\prime},y_{m+1})d\mathbf{y}^{\prime}\\
+\int_{0}^{\tau}\int_{\mathbb{R}^{m}}\Gamma(\mathbf{y},\mathbf{y}^{\prime},\tau,\tau^{\prime})\omega^{\prime}(\mathbf{y}^{\prime},y_{m+1},\tau^{\prime})d\mathbf{y}^{\prime}d\tau^{\prime}\text{ on }\mathbb{R}^{m}\times\left(0,\Delta\right)\label{eq:Preservation-Convolution}
\end{multline}
where $\Gamma$ is the fundamental solution for $\mathcal{L}^{\prime}$
(whose construction was first detailed by \cite{levi1907sulle}).
We first note that \ref{itm:Preservation-Continuous} follows immediately
if $\varphi$ is convex, as shown below.
\begin{lemma}
If $\varphi$ is convex w.r.t. the order $\leq_{\mathcal{A}}$ (see
Definition \emph{\ref{def:ControlReduction-Convex}}), then for all $x_{m+1}$,
$\mathbf{y} \mapsto \varphi^\prime(\mathbf{y};x_{m+1})$ is continuous
on $\mathbb{R}^{m}$.
\end{lemma}
\begin{proof}
We have assumed that $\varphi\equiv\varphi(\mathbf{x},x_{m+1})$
is convex w.r.t. $\leq_{\mathcal{A}}$ on $\Omega$. From this it
follows that for all $x_{m+1}\in\Omega_{2}$, $\varphi$ is convex
in $\mathbf{x}$ on $\Omega_{1}\equiv(0,\infty)^{m}$ w.r.t.
to the order $\leq$ on $\mathbb{R}^{m}$. This in turn yields that
for all $x_{m+1}\in\Omega_{2}$, $\varphi$ is continuous in $\mathbf{x}$
on $\Omega_1$. 
Therefore, $\varphi^{\prime}\equiv\varphi^{\prime}(\mathbf{y};x_{m+1})$
is continuous in $\mathbf{y}$ on $\mathbb{R}^{m}$.\end{proof}
\begin{theorem}
Suppose \emph{\ref{itm:Preservation-Independence}}, \emph{\ref{itm:Preservation-UniformlyParabolic}}
and \emph{\ref{itm:Preservation-Regularity}}---\emph{\ref{itm:Preservation-Growth}}.
Suppose that $\varphi$ is CM w.r.t. the order $\leq_{\mathcal{A}}$
(see Definition \emph{\ref{def:ControlReduction-Convex}} and Lemma \emph{\ref{lem:ControlReduction-ConvexComposition}}).
Suppose further that for all $t\in(t_{n},t_{n+1}]$, $\omega$
is CM in $(\mathbf{x},x_{m+1})$ on $\Omega$
w.r.t. the order $\leq_{\mathcal{A}}$. Then, for all $t\in(t_{n},t_{n+1}]$,
$V$ is CM in $(\mathbf{x},x_{m+1})$
on $\Omega$ w.r.t. the order $\leq_{\mathcal{A}}$. In particular,
$V_{n}^{+}\equiv\lim_{t\downarrow t_{n}}V(\cdot,t)$ is CM provided this
limit exists.\end{theorem}
\begin{proof}
$\Gamma$ appearing in (\ref{eq:Preservation-Convolution}) depends
on $\mathbf{y}^{\prime}$ and $\mathbf{y}$ through $\mathbf{y}^{\prime}-\mathbf{y}$
alone since by \ref{itm:Preservation-Independence}, $a_{i,j}^{\prime}$,
$b_{i}^{\prime}$ and $c^{\prime}$ are independent of the spatial
variables \cite[Chap. 9: Thm. 1]{friedmanpartial}. Therefore
\begin{multline*}
u(\mathbf{y},y_{m+1},\tau)=\int_{\mathbb{R}^{m}}\Gamma(\mathbf{y}^{\prime}-\mathbf{y},\tau,0)\varphi^{\prime}(\mathbf{y}^{\prime},y_{m+1})d\mathbf{y}^{\prime}\\
+\int_{0}^{\tau}\int_{\mathbb{R}^{m}}\Gamma(\mathbf{y}^{\prime}-\mathbf{y},\tau,\tau^{\prime})\omega^{\prime}(\mathbf{y}^{\prime},y_{m+1},\tau^{\prime})d\mathbf{y}^{\prime}d\tau^{\prime}\text{ on }\mathbb{R}^{m}\times\left(0,\Delta\right).
\end{multline*}
Let $\log\mathbf{x}\equiv(\log x_{1},\ldots,\log x_{m})$.
From the above, whenever $x_{i}>0$ for all $i\leq m$,
\begin{multline*}
V(\mathbf{x},x_{m+1},t)=\int_{\mathbb{R}^{m}}\Gamma(\mathbf{y}^{\prime}-\log\mathbf{x},t_{n+1}-t,0)\varphi(e^{\mathbf{y}^{\prime}},x_{m+1})d\mathbf{y}^{\prime}\\
+\int_{0}^{t_{n+1}-t}\int_{\mathbb{R}^{m}}\Gamma(\mathbf{y}^{\prime}-\log\mathbf{x},t_{n+1}-t,\tau^{\prime})\omega(e^{\mathbf{y}^{\prime}},x_{m+1},t_{n+1}-\tau^{\prime})d\mathbf{y}^{\prime}d\tau^{\prime}\text{ on }\Omega\times\left(t_{n},t_{n+1}\right).
\end{multline*}
Denote by $\mathbf{x}\circ\mathbf{x}^{\prime}\equiv(x_{1}x_{1}^{\prime},\ldots,x_{m}x_{m}^{\prime})$
the elementwise product of $\mathbf{x}$ and $\mathbf{x}^{\prime}$.
The substitution $\mathbf{y}^{\prime}=\log(\mathbf{x}\circ\mathbf{x}^{\prime})$
into the above yields
\begin{multline*}
V(\mathbf{x},x_{m+1},t)=\int_{0}^{\infty}\ldots\int_{0}^{\infty}\Gamma(\log\mathbf{x}^{\prime},t_{n+1}-t,0)\varphi(\mathbf{x}\circ\mathbf{x}^{\prime},x_{m+1})\frac{1}{\prod_{i}x_{i}^{\prime}}d\mathbf{x}^{\prime}\\
+\int_{0}^{t_{n+1}-t}\int_{0}^{\infty}\ldots\int_{0}^{\infty}\Gamma(\log\mathbf{x}^{\prime},t_{n+1}-t,\tau^{\prime})\omega(\mathbf{x}\circ\mathbf{x}^{\prime},x_{m+1},t_{n+1}-\tau^{\prime})\frac{1}{\prod_{i}x_{i}^{\prime}}d\mathbf{x}^{\prime}d\tau^{\prime}\text{ on }\Omega\times\left(t_{n},t_{n+1}\right).
\end{multline*}
Since $\Gamma$ is $>0$ \cite[Chap. 2: Thm. 11]{friedmanpartial} (a
related, arguably more general result is given in \cite[Chap. IV: Prop. 1.11]{garroni1992green}),
from the convexity and monotonicity of $\varphi$ and $\omega$,
it follows immediately that $V(\mathbf{x},x_{m+1},t)$
is CM on $\Omega$ 
for any $t\in(t_{n},t_{n+1})$. 
\end{proof}
\begin{remark}
We can extend our construction
to $\overline{\Omega_1} \times \Omega_2$ by taking limits of $V$ up to the
boundary. Since the codomain of $V$ is Hausdorff, this extension is unique.
\end{remark}

\section{Commutativity of union and supremum}

Let $T$ be a poset with order $\leq$ satisfying the least-upper-bound
property. All supremums are taken w.r.t. $T$.
\begin{lemma}
\label{lem:Commutativity-ResultNonEmpty}Let $\mathcal{S}\equiv\{ S_{\alpha}\} _{\alpha\in\mathcal{A}}$
be an indexed family of nonempty subsets of $T$. Let $S\equiv\bigcup_{\alpha\in\mathcal{A}}S_{\alpha}$
and
\[
U \equiv \left\{ \sup S_\alpha \right\}_{\alpha \in \mathcal{A}} .
\]
Suppose $\mathcal{A}$ is nonempty.
Then, $\sup S=\sup U$ whenever $S$ is bounded above.\end{lemma}
\begin{proof}
Since $S$ is bounded above, its supremum $u$ must occur in $T$.
For each $\alpha$, $u$ is an upper bound of $S_{\alpha}$, and since
$S_{\alpha}$ is a nonempty subset of $T$, $\sup S_{\alpha}=u_{\alpha}$
for some $u_{\alpha}\in T$. Thus, $U=\{ u_{\alpha}\} _{\alpha\in\mathcal{A}}\subset T$.
Since $u_{\alpha}\leq u$ for each $\alpha$, $u$ is an upper bound
of $U$. Since $\mathcal{A}$ is nonempty, $U$ is nonempty and hence
$U$ has a least upper bound $u^{\prime}\in T$ with $u^{\prime}\leq u$.
Let $x\in S$. Then $x\in S_{\beta}$ for some $\beta$, and hence
$x\leq u_{\beta}\leq u^{\prime}$ so that $u^{\prime}$ is an upper
bound of $S$. Since $\sup S=u$, $u\leq u^{\prime}$ and hence $u^{\prime}=u$.
\end{proof}

\bibliographystyle{plainnat}
\bibliography{main}

\begin{thebibliography}{23}
\providecommand{\natexlab}[1]{#1}
\providecommand{\url}[1]{\texttt{#1}}
\expandafter\ifx\csname urlstyle\endcsname\relax
  \providecommand{\doi}[1]{doi: #1}\else
  \providecommand{\doi}{doi: \begingroup \urlstyle{rm}\Url}\fi

\bibitem[Azimzadeh et~al.(2014)Azimzadeh, Forsyth, and
  Vetzal]{azimzadeh2013hedging}
P.~Azimzadeh, P.~A. Forsyth, and K.~R. Vetzal.
\newblock Hedging costs for variable annuities under regime-switching.
\newblock In R.~S. Mamon and R.~J. Elliott, editors, \emph{Hidden Markov Models
  in Finance: Further Developments and Applications, Volume II}, volume 209 of
  \emph{International Series in Operations Research \& Management Science},
  pages 133--166. Springer, New York, NY, 2014.
\newblock \doi{10.1007/978-1-4899-7442-6_6}.

\bibitem[Bacinello et~al.(2011)Bacinello, Millossovich, Olivieri, and
  Pitacco]{bacinello2011variable}
A.~R. Bacinello, P.~Millossovich, A.~Olivieri, and E.~Pitacco.
\newblock Variable annuities: A unifying valuation approach.
\newblock \emph{Insurance Math. Econom.}, 49\penalty0 (3):\penalty0 285--297,
  2011.
\newblock \doi{10.1016/j.insmatheco.2011.05.003}.

\bibitem[Bauer et~al.(2008)Bauer, Kling, and Russ]{bauer2008universal}
D.~Bauer, A.~Kling, and J.~Russ.
\newblock A universal pricing framework for guaranteed minimum benefits in
  variable annuities.
\newblock \emph{ASTIN Bulletin}, 38\penalty0 (2):\penalty0 621--651, 2008.
\newblock \doi{10.2143/AST.38.2.2033356}.

\bibitem[Bergman et~al.(1996)Bergman, Grundy, and Wiener]{bergman1996general}
Y.~Z. Bergman, B.~D. Grundy, and Z.~Wiener.
\newblock General properties of option prices.
\newblock \emph{J. Finance}, 51\penalty0 (5):\penalty0 1573--1610, 1996.
\newblock \doi{10.1111/j.1540-6261.1996.tb05218.x}.

\bibitem[Bian and Guan(2008)]{bian2008convexity}
B.~Bian and P.~Guan.
\newblock Convexity preserving for fully nonlinear parabolic
  integro-differential equations.
\newblock \emph{Methods Appl. Anal.}, 15\penalty0 (1):\penalty0 39--52, 2008.
\newblock \doi{10.4310/MAA.2008.v15.n1.a5}.

\bibitem[Butrica et~al.(2009)Butrica, Iams, Smith, and
  Toder]{butrica2009disappearing}
B.~A. Butrica, H.~M. Iams, K.~E. Smith, and E.~J. Toder.
\newblock The disappearing defined benefit pension and its potential impact on
  the retirement incomes of baby boomers.
\newblock \emph{Social Security Bulletin}, 69\penalty0 (3):\penalty0 1--27,
  2009.

\bibitem[Chen and Forsyth(2008)]{chen08a}
Z.~Chen and P.~A. Forsyth.
\newblock A numerical scheme for the impulse control formulation for pricing
  variable annuities with a guaranteed minimum withdrawal benefit ({GMWB}).
\newblock \emph{Numer. Math.}, 109\penalty0 (4):\penalty0 535--569, 2008.
\newblock \doi{10.1007/s00211-008-0152-z}.

\bibitem[Chen et~al.(2008)Chen, Vetzal, and Forsyth]{chen2008effect}
Z.~Chen, K.~R. Vetzal, and P.~A. Forsyth.
\newblock The effect of modelling parameters on the value of {GMWB} guarantees.
\newblock \emph{Insurance Math. Econom.}, 43\penalty0 (1):\penalty0 165--173,
  2008.
\newblock \doi{10.1016/j.insmatheco.2008.04.003}.

\bibitem[Dai et~al.(2008)Dai, Kwok, and Zong]{dai2008guaranteed}
M.~Dai, Y.~K. Kwok, and J.~Zong.
\newblock Guaranteed minimum withdrawal benefit in variable annuities.
\newblock \emph{Math. Finance}, 18\penalty0 (4):\penalty0 595--611, 2008.
\newblock \doi{10.1111/j.1467-9965.2008.00349.x}.

\bibitem[Forsyth and Vetzal(2014)]{forsyth2013risk}
P.~A. Forsyth and K.~R. Vetzal.
\newblock An optimal stochastic control framework for determining the cost of
  hedging of variable annuities.
\newblock \emph{J. Econom. Dynam. Control}, 44:\penalty0 29--53, 2014.
\newblock \doi{10.1016/j.jedc.2014.04.005}.

\bibitem[Friedman(1964)]{friedmanpartial}
Avner Friedman.
\newblock \emph{Partial differential equations of parabolic type}.
\newblock Prentice-Hall, Englewood Cliffs, NJ, 1964.

\bibitem[Garroni and Menaldi(1992)]{garroni1992green}
M.~G. Garroni and J.~L. Menaldi.
\newblock \emph{Green functions for second order parabolic integro-differential
  problems}.
\newblock Number 275 in Pitman Research Notes in Mathematics Series. Longman
  Scientific \& Technical, Harlow, UK, 1992.

\bibitem[Holz et~al.(2012)Holz, Kling, and Ru{\ss}]{holz2012gmwb}
D.~Holz, A.~Kling, and J.~Ru{\ss}.
\newblock {GMWB} for life an analysis of lifelong withdrawal guarantees.
\newblock \emph{Zeitschrift f{\"u}r die gesamte Versicherungswissenschaft},
  101\penalty0 (3):\penalty0 305--325, 2012.
\newblock \doi{10.1007/s12297-012-0193-3}.

\bibitem[Huang and Forsyth(2012)]{huang:2010}
Y.~Huang and P.~A. Forsyth.
\newblock Analysis of a penalty method for pricing a guaranteed minimum
  withdrawal benefit ({GMWB}).
\newblock \emph{IMA J. Numer. Anal.}, 32\penalty0 (1):\penalty0 320--351, 2012.
\newblock \doi{10.1093/imanum/drq044}.

\bibitem[Huang and Kwok(2014)]{huang2013analysis}
Y.~T. Huang and Y.~K. Kwok.
\newblock Analysis of optimal dynamic withdrawal policies in withdrawal
  guarantee products.
\newblock \emph{J. Econom. Dynam. Control}, 45:\penalty0 19--43, 2014.
\newblock \doi{10.1016/j.jedc.2014.05.008}.

\bibitem[Janson and Tysk(2004)]{janson2004preservation}
S.~Janson and J.~Tysk.
\newblock Preservation of convexity of solutions to parabolic equations.
\newblock \emph{J. Differential Equations}, 206\penalty0 (1):\penalty0
  182--226, 2004.
\newblock \doi{10.1016/j.jde.2004.07.016}.

\bibitem[Levi(1907)]{levi1907sulle}
Eugenio~Elia Levi.
\newblock Sulle equazioni lineari totalmente ellittiche alle derivate parziali.
\newblock \emph{Rendiconti del Circolo Matematico di Palermo}, 24\penalty0
  (1):\penalty0 275--317, 1907.
\newblock \doi{10.1007/BF03015067}.

\bibitem[Milevsky and Salisbury(2006)]{milevsky2006financial}
M.~A. Milevsky and T.~S. Salisbury.
\newblock Financial valuation of guaranteed minimum withdrawal benefits.
\newblock \emph{Insurance Math. Econom.}, 38\penalty0 (1):\penalty0 21--38,
  2006.
\newblock \doi{10.1016/j.insmatheco.2005.06.012}.

\bibitem[Ngai and Sherris(2011)]{ngai2011longevity}
A.~Ngai and M.~Sherris.
\newblock Longevity risk management for life and variable annuities: The
  effectiveness of static hedging using longevity bonds and derivatives.
\newblock \emph{Insurance Math. Econom.}, 49\penalty0 (1):\penalty0 100--114,
  2011.
\newblock \doi{10.1016/j.insmatheco.2011.02.009}.

\bibitem[Pasdika and Wolff(2005)]{pasdika2005coping}
U.~Pasdika and J.~Wolff.
\newblock Coping with longevity: The new {German} annuity valuation table {DAV}
  2004 {R}.
\newblock In \emph{Proceedings of the Living to 100 and Beyond Symposium},
  Orlando, FL, 2005. Society of Actuaries.

\bibitem[Piscopo and Haberman(2011)]{piscopo2011valuation}
G.~Piscopo and S.~Haberman.
\newblock The valuation of guaranteed lifelong withdrawal benefit options in
  variable annuity contracts and the impact of mortality risk.
\newblock \emph{North American Actuarial Journal}, 15\penalty0 (1):\penalty0
  59--76, 2011.
\newblock \doi{10.1080/10920277.2011.10597609}.

\bibitem[Rockafellar(1970)]{rockafellar1997convex}
R.~T. Rockafellar.
\newblock \emph{Convex analysis}.
\newblock Princeton University Press, Princeton, NJ, 1970.
\newblock \doi{10.1515/9781400873173}.

\bibitem[Wang and Forsyth(2008)]{wang2008maximal}
J.~Wang and P.~A. Forsyth.
\newblock Maximal use of central differencing for {Hamilton-Jacobi-Bellman
  PDEs} in finance.
\newblock \emph{SIAM J. Numer. Anal.}, 46\penalty0 (3):\penalty0 1580--1601,
  2008.
\newblock \doi{10.1137/060675186}.

\end{thebibliography}

\end{document}